\documentclass[iop]{emulateapj}

\newcommand{\spitzer}{\mbox {\it Spitzer}}
\newcommand{\msx}{{\it MSX}}
\newcommand{\um}{$\mu$m}
\newcommand{\hii}{\mbox{\ion{H}{2}~}}
\renewcommand{\deg}{$^\circ$}
\newcommand{\degree}{^{\circ}}
\newcommand{\Msun}{M$_{\sun}$}
\newcommand{\Lsun}{L$_{\sun}$}
\newcommand{\kms}{km~s$^{-1}$}

\slugcomment{For Publication in the CCCP Special Issue of the Astrophysical Journal Supplement Series}

\shorttitle{A Pan-Carina YSO Catalog}
\shortauthors{Povich et al.}

\begin{document}

\title{A Pan-Carina YSO Catalog: Intermediate-Mass Young Stellar
  Objects in the Carina Nebula Identified Via Mid-Infrared Excess Emission}

\author{Matthew S. Povich,\altaffilmark{1,2}
  Nathan Smith,\altaffilmark{3}
  Steven R. Majewski,\altaffilmark{4} 
  Konstantin V. Getman,\altaffilmark{1}
  Leisa K. Townsley,\altaffilmark{1} 
  Brian L. Babler,\altaffilmark{5} 
  Patrick S. Broos,\altaffilmark{1} 
  R\'{e}my Indebetouw,\altaffilmark{4} 
  Marilyn R. Meade,\altaffilmark{5}
  Thomas P. Robitaille,\altaffilmark{6,7}
  Keivan G. Stassun,\altaffilmark{8,9}
  Barbara A. Whitney,\altaffilmark{5,10}
  Yoshinori Yonekura,\altaffilmark{11,12}
  \& Yasuo Fukui\altaffilmark{13}
}

\altaffiltext{1}{Department of Astronomy and Astrophysics, The Pennsylvania
  State University, 525 Davey Laboratory, University Park, PA 16802}
\altaffiltext{2}{NSF Astronomy and Astrophysics Postdoctoral Fellow; povich@astro.psu.edu}
\altaffiltext{3}{Steward Observatory, University of Arizona, 933 North Cherry Avenue, Tucson, AZ 85721}
\altaffiltext{4}{Department of Astronomy, University of Virginia, P. O. Box 400325, Charlottesville, VA 22904-4325}
\altaffiltext{5}{Department of Astronomy, University of Wisconsin, 475
N. Charter Street, Madison, WI 53706}
\altaffiltext{6}{Harvard--Smithsonian Center for Astrophysics, 60 Garden Street, Cambridge, MA 02138}
\altaffiltext{7}{{\it Spitzer} Postdoctoral Fellow}
\altaffiltext{8}{Department of Physics and Astronomy, Vanderbilt University, Nashville, TN 37235} 
\altaffiltext{9}{Department of Physics, Fisk University, 1000 17th Avenue N., Nashville, TN 37208}
\altaffiltext{10}{Space Science Institute, 3100 Marine Street, Suite A353, Boulder, CO 80303}
\altaffiltext{11}{Department of Physical Science, Osaka Prefecture University, 1-1 Gakuen-cho, Sakai, Osaka 599-8531, Japan}
\altaffiltext{12}{Center for Astronomy, Ibaraki University, 2-1-1 Bunkyo, Mito, Ibaraki 310-8512, Japan}
\altaffiltext{13}{Department of Astrophysics, Nagoya University, Furo-cho, Chikusa-ku, Nagoya 464-8602, Japan}

\begin{abstract}
We present a catalog of 1439 young stellar objects (YSOs) spanning the
1.42 deg$^2$ field surveyed by the {\it Chandra} Carina Complex
Project (CCCP), which includes the major ionizing clusters and the
most active sites of ongoing star formation within the Great Nebula in
Carina. Candidate YSOs were identified via infrared (IR) excess
emission from dusty circumstellar disks and envelopes, using data from
the {\it Spitzer Space Telescope} 
(the Vela--Carina survey) and the Two-Micron All Sky Survey. We model
the 1--24~\um\ IR spectral energy distributions of the YSOs to
constrain physical properties.  Our Pan-Carina YSO Catalog (PCYC) is
dominated by intermediate-mass (2~\Msun$<m\la 10$~\Msun) objects with
disks, including Herbig Ae/Be stars and their less evolved
progenitors.  The PCYC provides a valuable complementary dataset to
the CCCP X-ray source catalogs, identifying 1029 YSOs in Carina with
no X-ray detection.  We also catalog 410 YSOs with X-ray counterparts,
including 62 candidate protostars. Candidate protostars with X-ray
detections tend to be more evolved than those without.  In most cases,
X-ray emission apparently originating from intermediate-mass,
disk-dominated YSOs is consistent with the presence of low-mass
companions, but we also find that X-ray emission correlates with
cooler stellar photospheres and higher disk masses. We suggest that
intermediate-mass YSOs produce X-rays during their early pre-main
sequence evolution, perhaps driven by magnetic dynamo activity during
the convective atmosphere phase, but this emission dies off as the
stars approach the main sequence. Extrapolating over the stellar
initial mass function scaled to the PCYC population, we predict a
total population of ${>}2\times 10^4$ YSOs and a present-day star
formation rate (SFR) of ${>}0.008$~\Msun~yr$^{-1}$.  The global SFR in
the Carina Nebula, averaged over the past ${\sim}5$~Myr, has been
approximately constant.
\end{abstract}


\section{Introduction}


The Great Nebula in Carina (NGC 3372) is among the most active
massive star-forming complexes in the Galaxy. Over 70 O-type stars
ionize a giant \hii\ region \citep{NS06a} 
and power a bipolar superbubble that has erupted from a 100-pc scale
giant molecular cloud 
complex \citep{NS00}. 
The Carina Nebula is often considered to be the
nearest analog of extragalactic starburst regions
\citep[$d=2.3$~kpc;][]{NS06b}, 
and it is also the least obscured. Carina thus presents a rich
target for the study of resolved young stellar populations evolving in
an environment dominated by some of the most massive stars
known. While numerous studies have targeted individual rich ionizing
clusters in Carina \citep{FMF80,MJ93,DGE01,JA07,HS10}, 
global observations of the stellar populations
across the complex are challenging. Due to its large angular size
(${>}2\degree\times 
1.5\degree$), it is difficult to achieve the requisite wide
observational coverage of the Carina Nebula at high spatial resolution. Its position
near the Galactic midplane ($b=-0\fdg6$) means that wide-field images at visual
and infrared (IR) wavelengths are dominated by contaminating
populations of unassociated stars. The {\it Chandra} Carina Complex
Project (CCCP), a 1.2 Ms total X-ray integration covering a 1.42 deg$^{2}$
field in the center of the Carina 
Nebula, provides an unprecedented opportunity to overcome these
observational challenges and study the young stellar populations from
a global perspective, yet at the detailed level of resolving individual stars
\citep{CCCPintro}. 

X-ray observations of star-forming regions are highly sensitive to
low-mass pre-main-sequence (PMS) stars (T Tauri stars), which generate
hard X-rays via convectively-driven magnetic flaring activity
\citep{TP05} 
and to OB stars that produce soft
X-rays primarily through micro-shocks in their strong stellar winds \citep{LW80,FPP87}.
%
%
Both classical T Tauri stars with circumstellar disks and
protostars with accreting 
envelopes are observed to be significantly less luminous in
X-rays compared to diskless, weak-lined T Tauri stars
\citep{KS04,XEST,COUPprotostars}. 
The CCCP observations are expected to be insensitive to
intermediate-mass (2~\Msun$<m<10$~\Msun) 
young stellar objects (YSOs), including Herbig 
Ae/Be stars with disks and their less evolved progenitors, which lack
a known mechanism for the production of strong X-ray emission
\citep{BS06}. This
important stellar population traces recent and ongoing
star formation in massive complexes like Carina. Because such stars
have high bolometric luminosities and possess
dusty disks and envelopes that reprocess the stellar radiation into
strong mid-IR emission,
surveys with the {\it Spitzer Space Telescope} are {\it very}
sensitive to them \citep[][hereafter PW10]{XK08,R08,MK09,PW10}.
The entire Carina Nebula was included in the {\it Spitzer} Vela--Carina
survey.
This complementarity between the IR and
X-ray observations motivates us to search for candidate YSOs
throughout the Carina Nebula using Vela--Carina survey data in
conjunction with supplementary IR datasets.

The Carina complex contains at least
 $6.7\times 10^5$~\Msun\ in molecular gas mass \citep{DG88} but
 hosts few bright embedded IR sources or other signposts of
 ongoing massive star formation; hence historically the Carina Nebula
 has been regarded as a relatively evolved \hii\ region, with a
 current star formation rate (SFR) that is negligible in comparison
 with its historical rate \citep[][and
 references therein]{SB08}. Within the past decade, a new view of star
 formation in Carina has emerged, in which active
 star formation has migrated to the outer reaches of the nebula, in
 particular the region known as the South Pillars, where remnant
 molecular cloud material has been shredded and sculpted into myriad
 clumps, pillars, and elephant-trunk structures by stellar winds and
 radiation, the two dominant feedback mechanisms produced by massive
 stars during their lifetimes \citep{TM96,NS00,JR04,SB07}. \citet[][hereafter 
 SP10]{spitzcar} presented the first \spitzer\ imaging study of Carina
 using the Infrared Array Camera \citep[IRAC;][]{IRAC} and found over
 900 candidate YSOs widely distributed throughout two large fields
 covering the South Pillars and a portion of the opposite wall of the
 southern superbubble lobe. Carina remains a highly active
 star-forming region, but it may be unusual among Galactic giant
 molecular glouds (GMCs)
 in the sense that the dominant mode of current star formation activity
 appears to be triggering by feedback from the local massive
 stellar population (SP10). 

Our present study expands upon the \spitzer\ point-source
results presented by SP10 in several ways: (1) The Vela--Carina
observations extend \spitzer\ coverage to the entire Carina Nebula,
including the central 
ionizing clusters and regions to the North. The wide-field coverage
enables us to analyze relatively empty ``control'' fields far from the
Carina GMC and correct for contamination in our YSO sample from
unassociated sources with similar mid-IR colors. (2) We incorporate
24~\um\ photometry from the Multiband Imaging Photometer for \spitzer\
\citep[MIPS;][]{MIPS} to constrain the evolutionary stages and
luminosities of individual YSOs better than what is possible with IRAC
photometry 
alone. (3) We analyze the subsample of YSOs that match CCCP catalog sources to
investigate the physical properties of candidate X-ray-emitting protostars and
intermediate-mass PMS stars with disks. 




In this paper, we 
present a highly-reliable catalog of
1,439 predominantly intermediate-mass YSOs throughout the Carina
Nebula, selected via their IR excess emission. This Pan-Carina YSO Catalog 
(PCYC) complements and augments the CCCP X-ray catalog
\citep{CCCPcatalog}. The data analysis steps used to
construct the PCYC are summarized in \S2, and the methodology used to
model the physical properties of the YSOs is described in \S3. 
In \S4, we present the
ensemble properties of the YSO population, in \S5 we analyze the
subset of our sample associated with X-ray emission, and in \S6 we
highlight five noteworthy YSO clusters and sub-groups and several
interesting individual YSOs. We discuss our results in the context of
the star formation history of the Carina Nebula and the physical
mechanism(s) powering X-ray emission from YSOs in \S7.
We summarize our conclusions in \S8.

\section{Data Analysis and YSO Sample Selection}

\subsection{Observations}
The Vela--Carina survey (\spitzer\ Proposal ID 40791, PI
S. R. Majewski), was modeled after 
the Galactic Legacy Infrared 
Mid-Plane Survey Extraordinaire \citep[GLIMPSE;][]{GLIMPSE}. It used
the 4 mid-IR bands of \spitzer/IRAC (3.6, 4.5, 5.8, and 8.0~\um) to
image 86 deg$^2$ of the Galactic plane in a strip 2\deg\ high in
latitude over longitudes $255\degree < l < 295\degree$ \citep{VCex},
an area than included the 
Carina complex. PSF-fitting photometry was performed using the GLIMPSE
point-source extractor, a modified version of DAOPHOT \citep{Daophot}
optimized for crowded fields with strong, varying background nebular
emission. The GLIMPSE extractor produces two source lists, the
99.5\%-reliable Point Source Catalog and the more complete Point
Source Archive.\footnote{For details of the data processing and
  products, go to \\
  \url[http://irsa.ipac.caltech.edu/data/SPITZER/GLIMPSE/doc/glimpse1_dataprod\v2.0.pdf]{http://irsa.ipac.caltech.edu/data/SPITZER/GLIMPSE/doc/glimpse1\_dataprod\_v2.0.pdf}.} 
The depth of the Vela--Carina photometry is
very similar to that of GLIMPSE, reaching $[3.6]\la 15.5$~mag (less
sensitive in regions of bright nebular emission). Both the Catalog and
the Archive sources have been matched (bandmerged) with $JHK_S$
photometry from the Two-Micron All-Sky Survey Point Source Catalog
\citep[2MASS;][]{2MASS}, and the resulting source lists provide
photometry in 7 IR bands, from 1.2~\um\ through 8.0~\um. We use
primarily the highly-reliable Catalog for this work, as experience
shows that this drastically reduces the number of false-positives,
candidate YSOs that are generated as the result of systematic
photometric errors. The Vela--Carina Catalog contains 60,515 point
sources inside the CCCP survey area (Table~\ref{numbers}).
We do retrieve a few
sources from the Archive that are matched to bright candidate YSOs
detected by the {\it Midcourse Space Experiment} (\msx), as described
in \S\ref{MSX} below.


\begin{deluxetable*}{lrrl}
\tabletypesize{\scriptsize}
\tablecaption{Summary of Vela-Carina Point Sources in CCCP Field\label{numbers}}
\tablewidth{0pt}
\tablehead{
  \colhead{Sources} & \colhead{All} & \colhead{CCCP-matched} & \colhead{Description}
}
\startdata
In Vela--Carina Catalog & 60,515 & 4,664 & Detected IRAC point sources, 99.5\% reliable \\
Fit with SED Models     & 54,155 & 4,080 & Detected in ${\ge}4$ of 7 2MASS+IRAC bands \\
Stellar Photospheres    & 50,586 & 3,444 & Well-fit by stellar atmosphere SEDs \\
Possible IR Excess      & 3,569 & 636 & Poorly-fit by stellar atmosphere SEDs \\
Marginal IR Excess      & 1,923 & 213 & Excess in only IRAC [5.8] or [8.0] band \\
Reliable IR Excess      & 1,646 & 423 & Intrinsically red objects \\
\tableline
In PCYC\tablenotemark{a}          & 1,439 & 410 & YSO catalog, corrected for contamination \\
Stage 0/I               &  247 &  62 & Dominant SED models: infalling envelopes \\
Stage II                &  478 & 190 & Dominant SED models: optically thick disks \\
Stage III               &   83 &  13 & Dominant SED models: optically thin disks \\
Ambiguous               &  631 & 144 & Inconclusive SED model classification
\enddata
\tablenotetext{a}{Includes 3 Vela-Carina Archive sources with \msx\ counterparts.}
\end{deluxetable*}

The CCCP catalog contains over 14,000 point sources detected by the
Advanced CCD Imaging Spectrometer \citep[ACIS;][]{ACIS} of the {\it
  Chandra X-ray Observatory}.
\citet{CCCPcatalog} matched the Vela--Carina Archive to the CCCP
X-ray catalog and found 6,543 counterparts to CCCP
sources. The Vela--Carina Catalog is a subset of the Archive, and
provides mid-IR counterparts for 4,664 CCCP sources (Table~\ref{numbers}).

The Carina Nebula was observed
with {\it Spitzer}/MIPS during late July 2007 as part of program GO-30848
(MIPSCAR; P.I. N.\ Smith). The program consisted of a series of
MIPS scans that covered most of the area of the star forming region
but avoided the bright emission of the Homunculus nebula around
$\eta$ Carinae. This yielded a large mosaic image of the 24 $\micron$
emission that contains the entire CCCP survey area. 

\subsection{Identification of Candidate YSOs}

%
YSOs possess dusty circumstellar disks and infalling envelopes that
reprocess radiation from the central stars, producing characteristic IR
excess emission. YSOs can be identified
in broad-band photometric imaging observations via their IR colors or
spectral energy distributions (SEDs).
Our procedure for identifying 
candidate YSOs is adapted from
\citet{P09}, SP10, and PW10. 
Here we outline the main steps used in constructing our
highly-reliable catalog of candidate YSOs, and we summarize the number
of sources found within the CCCP survey area for each step in Table~\ref{numbers}.
Using
a $\chi^2$-minimization SED fitting tool \citep{fitter}, we fit
reddened \citet{Kurucz} 
stellar atmospheres \citep[allowing $A_V$ to vary between 0 and 40~mag
according to the
extinction law of][]{I05} to the SEDs of
210,742 sources in the 
Vela-Carina Catalog located
within 
a $\Delta l\times \Delta b = 2\fdg8\times 2\degree$ field, 54,155 of
which fell within the CCCP survey area
(Table~\ref{numbers}).\footnote{The full field analyzed is 
  much larger than the CCCP field; regions of active star
  formation identified in this 
  wider field will be the subject of a future paper.}  
Only sources detected in
$N_{\rm data}\ge 4$
of the 7 Catalog bands
($\lambda=1$--8~\um) were fit. To avoid biasing the fits for
sources where the photometric uncertainties may have been
underestimated in the Catalog, before fitting any models we
conservatively reset the uncertainties to a minimum value of 10\%.  
We considered sources for which
$\chi_0^2/N_{\rm data}\le 2$ ($\chi_0^2$ is the
unreduced goodness-of-fit parameter of the best-fit model; SP10) to be
well-fit by stellar 
atmosphere SEDs; among sources in the CCCP field that were
fit with SED models, 93.4\% were consistent with
normally-reddened stellar photospheres (Table~\ref{numbers}) and hence
were removed from consideration for the PCYC. 

We then filtered out sources
with ``marginal'' IR excess emission, where the excess appears in only
the single IRAC [5.8] or [8.0] 
band, using the SP10 color
criteria, modified to de-redden the $[3.6]-[4.5]$ colors of background
stars viewed through the Carina cloud by up to $A_V=20$~mag. 
Using $[\lambda]$ to denote magnitudes in the various IRAC bands and
$\delta([\lambda_i]-[\lambda_j])$ for the uncertainties on the colors
computed from the (minimum 10\%) uncertainties on Catalog flux
densities, the selection criteria were as follows:
\begin{eqnarray*}
  [3.6]-[4.5]\phm{|} & > & \delta([3.6]-[4.5]) + E([3.6]-[4.5]) \\
{\rm OR}~~
\left|[4.5]-[5.8]\right| & > & \delta([4.5]-[5.8]) \\
{\rm AND}~~
  [5.8]-[8.0]\phm{|} & > & \delta([5.8]-[8.0]).
\end{eqnarray*}
The color excess used for the de-reddening was calculated as
\begin{displaymath}
  E([3.6]-[4.5]) = A_V\frac{(\kappa_{3.6} -
  \kappa_{4.5})}{\kappa_V} = 0.0135 A_V,
\end{displaymath}
where the $\kappa_{\lambda}$ are opacities given by the extinction law
\citep{I05}.
This step is crucial to distinguish intrinsically red objects from
sources affected by potential systematic photometric errors
\citep{P09}; IRAC [5.8] and 
[8.0] are less sensitive to point sources and more affected by nebular
emission compared to IRAC [3.6] and [4.5]. 
We discarded 1,923 of these
``marginal'' IR excess sources, leaving
1,646 ``reliable'' intrinsic IR excess sources, which form the basis for
our sample of candidate YSOs (Table~\ref{numbers}). Throwing out the
majority of the possible IR excess sources in this manner may seem
overly conservative, but we note that among the CCCP-matched
subsample, only 1/3 of sources were caught by this filter
(Table~\ref{numbers}). The 
Vela--Carina survey images are dominated by K- and M-type
giants in the Galactic field that are not expected to emit
X-rays \citep{RR95}. 
If the majority of marginal IR
excess sources were actually due to emission from disks, perhaps
transitional disks with large inner holes, we would
expect the fraction of sources rejected at this step among the X-ray
detected subsample to be similar to, if not higher than, the fraction rejected from
the full sample. Instead, 
the significantly lower frequency of marginal IR excess sources with X-ray
counterparts affirms that the majority of these
sources are unassociated field stars. The 213 CCCP-matched sources in
Table~\ref{numbers} rejected on 
account of marginal IR excess are examined again by
\citet[][hereafter P11]{Paper II}. 

\subsection{MIPS 24 \um\ Aperture Photometry}

Our data analysis is tailored to the goal of fitting YSO model SEDs from
\citet[][hereafter RW06]{grid} to the 
available IR photometry 
of each candidate YSO.
Because the SEDs of highly embedded
YSOs peak in the thermal IR near 100~\um, photometry at
$\lambda > 10$~\um\ is often required to constrain bolometric
luminosities and to distinguish
envelope-dominated (Stage 0/I)
from disk-dominated (Stage II) YSOs \citep[RW06;][]{I07}. 
Following PW10, we measured MIPS 24~\um\ flux densities using aperture
photometry. 
We located the position of each
candidate YSO and centered 
an extraction aperture of radius 3.5\arcsec\ 
and background annulus
of inner and outer radii 7\arcsec\ and 13\arcsec, respectively,
on the MIPS 24~\um\ mosaic of Carina.
We extracted 24~\um\ flux densities for each source, 
estimating the
background level using the Daophot MMM algorithm \citep{Daophot}. 
Our choice of extraction aperture required an aperture
correction\footnote{MIPS Instrument Handbook v1.0, \url[http://ssc.spitzer.caltech.edu/mips/mipsinstrumenthandbook/]{http://ssc.spitzer.caltech.edu/mips/mipsinstrumenthandbook/}}
of 2.8.
The uncertainty introduced by the aperture correction is more
than offset by the greater accuracy of the local background
determination and the ability to separate close sources (PW10). 

PW10 analyzed MIPSGAL 24~\um\ survey data \citep{MIPSGAL} covering the M17 SWex IR
dark cloud. The MIPSCAR observations are shallower than the
MIPSGAL observations and affected by extremely bright nebular emission
produced by warm dust in the Carina \hii region. While ${\sim}65\%$ of
the YSOs cataloged by PW10 were detected at 24~\um, the detection
fraction is only ${\sim}10\%$ for the PCYC. 
We can, however, derive useful upper limits on 24~\um\ flux densities in the
majority of cases. Upper limits provide critical constraints for the SED
fitting, since the RW06 models may predict variations of more than an
order of magnitude in 24~\um\ flux density even when the IRAC flux densities
are well-measured.
Detections and upper limits were defined as follows: The formal
uncertainty $\sigma$ on 
the aperture photometry is 
generally dominated by 
the uncertainty on the photometric background determination but also takes into
account photon counting statistics (including the contribution of
the zodiacal background light). 
Taking $F^{\prime}$ to be the extracted flux density, sources for
which $F^{\prime}_{24}<5\sigma$ were considered undetected at 24~\um, and
upper limits of $F_{24}<5\sigma+F^{\prime}_{24}$ for
$F^{\prime}_{24}>0$ and $F_{24}<5\sigma$ for
$F^{\prime}_{24}\le 0$ were assigned.
The 
results of our aperture photometry are presented in Table~\ref{obs},
in which the 24~\um\ quality flags (column 20) are set to 1 for
detections and 3 for upper limits.
No photometric information can be obtained for sources falling within
regions of the 24~\um\ mosaic where the nebular emission saturates
(for example, in the vicinity of $\eta$ Car); these are given quality flags of 0. 
For the 24~\um\ detections we set the minimum
uncertainty to $0.15F_{24}$ to account for systematic
uncertainties. 
One potential source of systematics is the mid-IR extinction law for
$\lambda>10$~\um, which 
appears to vary with 
environment.  
Recent observational work with {\it Spitzer} has found that, in
molecular clouds, the absorption at
24~\um\ is similar to, not less than, the absorption at 8~\um\
\citep[e.g.][]{KF07}. 
We therefore used the high-column-density ($A_K\ge 1$) mid-IR
extinction law of \citet{MM09} when fitting the
RW06 models to the SEDs of the candidate YSOs.

\subsection{\msx\ Detections of Luminous Candidate YSOs}\label{MSX}

We incorporate 8--21~\um\ photometry from the \msx\ Galactic Plane
Survey Point Source Catalog \citep{MSX} into our YSO
sample in 2 ways: (1) Spatial correlation of \msx\ sources with the
candidate YSOs and (2) visual identification of bright mid-IR sources 
within the CCCP field that were excluded from
the Vela-Carina Catalog because they were either marginally resolved
or saturated.  
Moderately saturated sources
were included in the more complete Vela-Carina Point Source Archive, and we
added 3 \msx\ sources 
with Archive matches to the sample.
Only 11 PCYC sources have \msx\ counterparts (indicated in
Table~\ref{obs}), but these include the most 
luminous YSOs in the cloud. Most of these objects were previously
cataloged by \citet{JR04} or \citet{SB07}. 
The addition of up to 4 \msx\ datapoints provides strong constraints on the mid-IR 
SEDs along with replacements for {\it Spitzer} photometry
suffering from saturation at 8.0 or 24~\um. 
 
\subsection{Removal of Contaminating IR Excess Sources}

 While it is dominated by YSOs, the sample of reliable IR excess sources
 (Table~\ref{numbers}) may also
 contain the following types of contaminants: variable
 stars, dusty asymptotic giant branch (AGB) stars, unresolved planetary
 nebulae, and background galaxies \citep{R08,P09}. 
Although few in number, luminous AGB stars are 
the most important contaminants, as they can
masquerade as massive YSOs. The majority of AGB stars have
$[8.0]-[24]<2.2$~mag, and the few extreme AGB stars that are redder
are highly luminous and would be conspicuously bright \citep{BW08,P09}. 
Applying this color cut removed 1\% of
sources. We
correct for
the remaining contamination by analyzing the 
spatial distribution of IR excess sources.

While YSOs associated with the Carina Nebula exhibit a high degree of
clustering (SP10),
contaminating sources, including foreground YSOs, are distributed
uniformly.
We used the nearest-neighbor group-finding algorithm described by
\citet{P09} to identify sources exhibiting a 
significant degree of clustering with respect to 3 
``control'' fields located outside of the CCCP
field. The algorithm was tuned to be sensitive to
groupings of ${\ge}10$ sources. The control fields were $0\fdg5\times 0\fdg5$
boxes centered at 
$(l,b)=(286\fdg4,-0\fdg9)$, $(286\fdg6,0\fdg2)$, and
$(288\fdg3,0\fdg1)$, containing a mean surface density of reliable IR
excess sources of $\Sigma_{\rm con}=200$ deg$^{-2}$. The surface
density of reliable IR excess sources inside the 1.42 deg$^{2}$ CCCP
field is $\Sigma_{\rm CCCP}=1150$~deg$^{-2}$ (Table~\ref{numbers}),
hence the expected contamination fraction from sources unassociated with the
Carina Nebula is $\Sigma_{\rm con}/\Sigma_{\rm CCCP}=17\%$. 
After removing sources from the sample that did not satisfy the
clustering criterion, the source density in the final 
PCYC is $\Sigma^{\rm corr}_{\rm CCCP}=1000$~deg$^{-2}$, indicating a
residual contamination level of $\Sigma^{\rm corr}_{\rm
  CCCP}-(\Sigma_{\rm CCCP}-\Sigma_{\rm con})=50$~deg$^{-2}$, or 5\%.

\section{Model-Based YSO Characterization and Classification}


Basic IR photometric data for the 1439 YSOs in the PCYC are presented
in Table~\ref{obs}. Using the \citet{fitter} SED fitting tool, we fit YSO
models from RW06 to the available photometry of each source. We
allowed the fitting tool to accept a range of distances of $2.3\pm
0.05$~kpc, guided by the well-established distance to $\eta$~Car from
expansion measurements of the Homunculus nebula \citep{NS06b} and by the
assumption that the depth of the Carina Nebula along the line of sight
is not greater than its ${\sim}50$~pc diameter observed
on the sky.
We define the set $i$ of well-fit models for each YSO according to
$\chi^2_i - \chi^2_0 \le 2N_{\rm data}$, and assign a
$\chi^2$--weighted, normalized probability
$P_i$ to each model \citep{P09}. This enables us to construct
probability distributions for the model parameters and thereby
constrain physical properties such as luminosity, mass, and accretion
rate. 
In general, the parameters are not normally distributed, so 
the physical properties for individual sources must be interpreted with
care.
We present results from the model fitting to the PCYC sources in
columns (2)--(8) of Table~\ref{mod}. For each YSO, we list
characteristic values of 3 parameters along with corresponding
uncertainties. The characteristic values for the stellar parameters
mass ($X = M_{\star}$) and bolometric luminosity ($X=L_{\rm bol}$) are
calculated as the probability-weighted means of the parameter distributions,
\begin{equation}\label{stellar}
  \langle X\rangle = \sum P_i X_i. 
\end{equation}
Circumstellar parameters in the RW06 models, such as envelope accretion rate
($Y=\dot{M}_{\rm env}$), span many orders of magnitude. We
define the characteristic circumstellar parameters as the medians of
the parameter distributions,
\begin{equation}\label{circumstellar}
  \langle Y\rangle = \mu_{1/2}(Y),
\end{equation}
because this approach is less affected by any extreme outlying values
returned by the fitting procedure.
For both stellar and circumstellar parameters, we compute the
$1\sigma$ uncertainty $\sigma(X,Y)$ as the standard deviation on the
characteristic parameter value $\langle X,Y\rangle$.

The YSO models are divided into evolutionary stages that
parallel the well-known empirical T Tauri classification system based
on IR colors or spectral indices: Stage 0/I YSOs are still embedded in
their infalling, natal envelopes; Stage II YSOs have
optically thick circumstellar disks; Stage III YSOs have optically
thin disks (RW06). We construct probability distributions of evolutionary
stage for each YSO and classify a YSO as Stage 0/I, II, or III if
$\sum P_i(\rm Stage)\ge 0.67$; we refer to the classification of
sources not meeting this criterion as
``Ambiguous'' \citep{P09}. 
We prefer the RW06 stage classification system to the traditional T Tauri
class system because (1) it makes the model-dependent
nature of the taxonomy explicit rather than implicit, (2) the
incorporation of Ambiguous 
classifications  acknowledges the limitations of imposing distinct
categories on a continuous evolutionary
sequence, and (3) a model-based system provides physical insight into
intermediate- and high-mass YSOs for which a classification based on
the empirical colors of low-mass T Tauri stars may not be meaningful.
The distribution of stage classifications across the PCYC is presented
in Table~\ref{numbers}, and the classification for each YSO is given
in Table~\ref{mod} (column 8).  

The 410 YSOs detected in X-rays are identified in
Table~\ref{mod} by CCCP catalog name (column 9). \citet[][hereafter
F11]{CCCPclusters} 
divided the CCCP field into 4 broad regions (A, B, C, and D)
and also identified 20 X-ray
clusters and 31 smaller groups. We assign every PCYC
source to one of these regions and possibly also to one of the
clusters/groups according to the region/cluster/group membership of the nearest CCCP
catalog source (columns 10 and 11 of Table~\ref{mod}).

The PCYC (Tables \ref{obs} and \ref{mod}) is a
compilation of individual candidate YSOs in Carina that we hope will
provide a rich dataset for future follow-up studies. For the present
work, we are concerned primarily with the ensemble properties of the
population. The large sample size
provides robust statistics to compensate for potentially large
uncertainties on individual YSOs whose physical
properties, such as stellar mass and bolometric luminosity, may not be
well-constrained by the RW06 models. 


\section{Global Properties of the Carina YSO Population}

\subsection{Spatial Distribution}\label{spatial}

\begin{figure*}
\epsscale{0.99}
\plotone{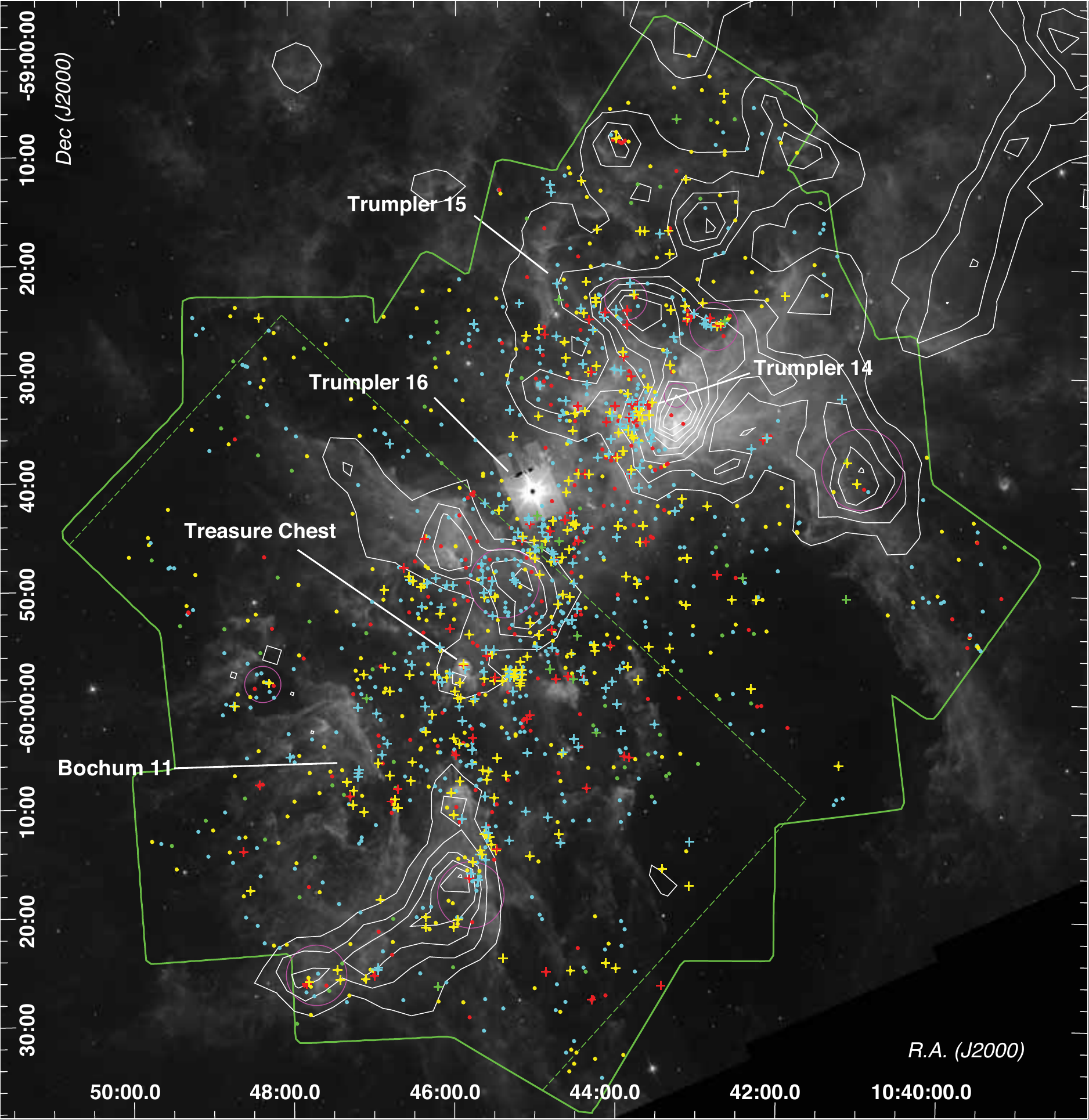}
\caption{\spitzer/IRAC 8.0~\um\ mosaic image with positions of the 1439 YSOs
  in the Pan-Carina YSO Catalog (PCYC) overplotted. Colors identify YSO evolutionary
  stages: red=Stage 0/I, yellow=Stage II,
    green=Stage III, cyan=ambiguous. YSOs with X-ray
  counterparts are marked by crosses. Contours are CO
  integrated intensity over the velocity range $V_{\rm LSR}=-30$~\kms\
  to $-10$~\kms, and the large magenta circles mark C$^{18}$O
  molecular cores \citep{YY05}.
  The solid green outline is the
  boundary of the CCCP survey area, while the dashed green lines
  enclose the area where the South Pillars observation of SP10
  overlaps the CCCP. The bright, saturated star near the image center is
  $\eta$ Car. Five well-known massive star clusters discussed in the text are labeled. 
\label{image}}
\end{figure*}

The positions of sources in the YSO catalog
are plotted on an IRAC
8.0~\um\ mosaic image from the Vela--Carina survey in
Figure~\ref{image}. While YSOs are distributed widely throughout the
CCCP field, the population generally traces the broad North--South
stellar ``backbone'' 
joining the major ionizing clusters Trumpler (Tr) 14 and 16 with the Collinder
(Cr) 228/South Pillars region \citep[][F11]{CCCPintro}. 
Contours of the major molecular clouds traced in CO emission and the
locations of 8 dense 
C$^{18}$O molecular cloud cores \citep{YY05} are also plotted in Figure~\ref{image}.
While the spatial distribution of YSOs throughout the Carina Nebula
exhibits highly complex structure, with clustering on multiple spatial
scales,
a general pattern is readily discernible. The majority of YSOs
are located inside \hii region cavities near,
but less frequently within, the boundaries of dense molecular clouds
and the ends of pillars. 

%
%
%
\begin{figure*}
\epsscale{0.99}
\plotone{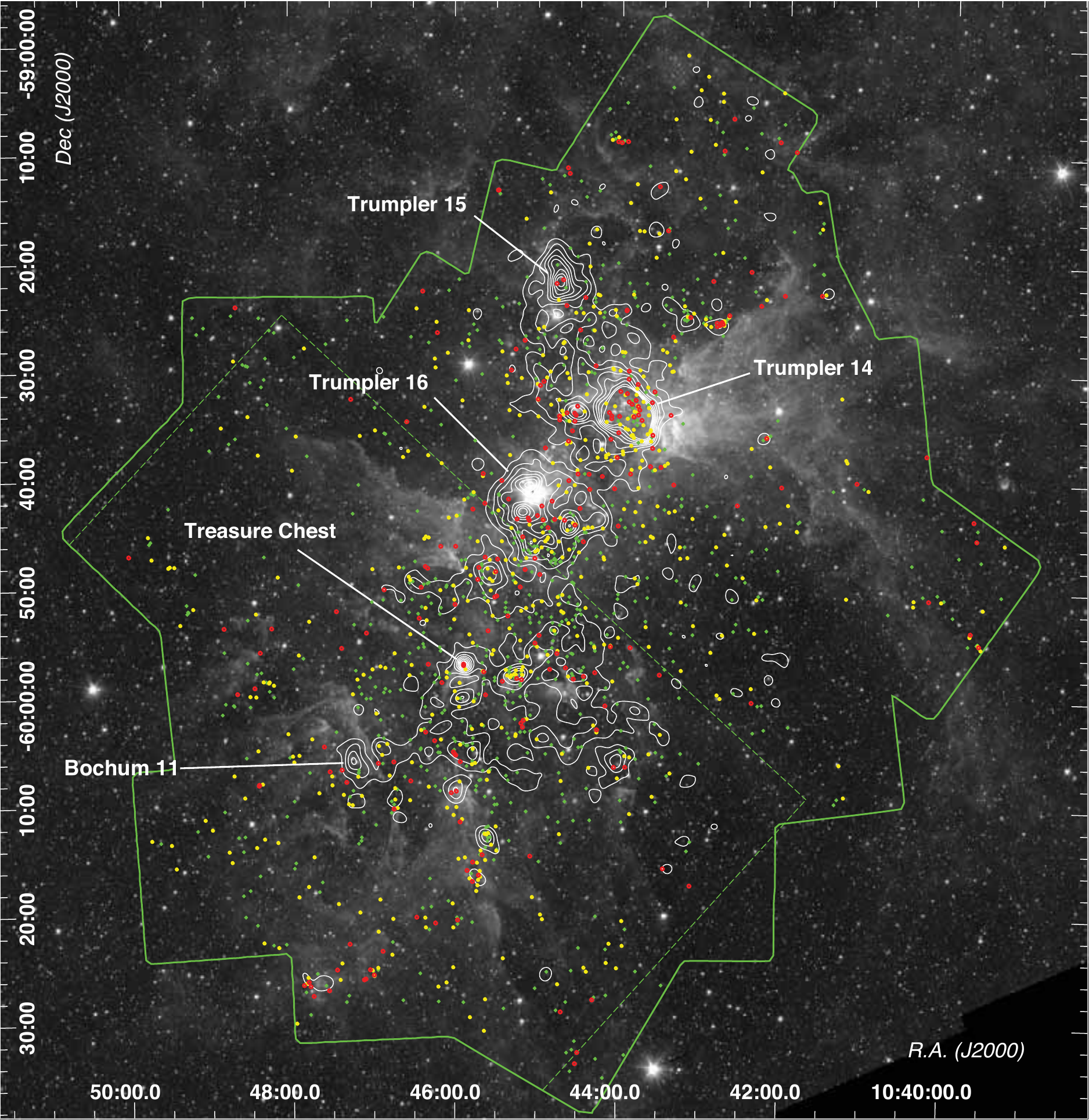}
\caption{\spitzer--IRAC 3.6~\um\ mosaic image with positions of the 1439 YSOs
  in the PCYC overplotted. Colors identify
  YSOs by mass range: red, $\langle M_{\star}\rangle\ge 3.1$~\Msun; yellow,
  2~\Msun$\le \langle M_{\star}\rangle<3.1$~\Msun; green, $\langle
  M_{\star}\rangle < 2$~\Msun. Contours are from F11 and
  show the surface density of X-ray sources. 
  The green outlines and cluster labels are the same as in Figure~\ref{image}.
\label{newimage}}
\end{figure*}

Another view of the PCYC is presented in  
Figure~\ref{newimage}, in which a Vela--Carina 3.6~\um\ mosaic is
presented with YSOs color-coded by mass. YSOs in 3 mass ranges roughly
corresponding to the mass ranges of Herbig Be stars, Herbig Ae stars,
and T Tauri stars are colored 
{\it red}, {\it yellow}, and {\it green}, respectively. Density
contours of CCCP sources from F11 depict all of the major clusters as
well as numerous 
smaller stellar groupings. 
The largest concentration of intermediate-mass YSOs is found not in the South
Pillars, but in Tr 14 itself. Numerous sub-clusters of Tr 16
(F11) also have associated YSOs. 

Several caveats regarding the \spitzer\ point-source detections must
be kept in mind when analyzing the PCYC. The strong and variable
nebular emission across the Carina Nebula, 
especially at 8.0 and 24~\um, creates complicated, position-dependent
variations in the sensitivity limit for mid-IR
point-source detections.
In addition, the IRAC images are
confusion-limited in the dense cluster centers. The PCYC is
very incomplete near $\eta$ Car, a highly saturated \spitzer\ source
visible at the center of 
Figures~\ref{image} and \ref{newimage}, and in the Treasure Chest,
a very young, embedded cluster \citep{TC},
where only 3 candidate 
YSOs are detected due to extreme crowding and nebulosity. Other
potential areas of sensitivity problems 
include the crowded center of Tr 14 and the ionization front
between Tr 14 and the Car I molecular cloud core to the west \citep{YY05,JA07} 
where
the diffuse mid-IR nebular emission is bright. Nevertheless, the
dearth of active star formation within the Car I cloud core appears to
be real, not simply a sensitivity effect. 

All of the \spitzer\ YSO sub-clusters cataloged by SP10 are apparent in
Figures~\ref{image} and \ref{newimage}. In addition, we detect 3 new,
tight clusters or groups of ${\sim}10$ 
YSOs in the Northwest corner of the CCCP area, 2 of which are
associated with significant overdensities of
X-ray sources (F11). 
We discuss these new YSO clusters in more detail in
\S\ref{butterflies}. 
YSOs are also detected embedded in
the ends of most dust pillars, both in the South Pillars and elsewhere
in the Carina Nebula. Some of these are the
driving sources of the 
Herbig-Haro jets found in {\it Hubble Space Telescope} images
\citep{HHjets}. YSOs are also associated with several of the candidate
proplyds identified by \citet{proplyds}, specifically proplyds
104405.4--592940 (PCYC 429), 104619.7--595044 (1126), 104632.9--600353
(1173), and 104519.3--594423 (842 and 841). All of the YSOs
associated with proplyds have either Stage 0/I or Ambiguous
classifications, and none 
are detected in X-rays.


\subsection{YSO Mass Function and Present-Day Star Formation Rate}\label{YMF_SFR}

\begin{figure}
\plotone{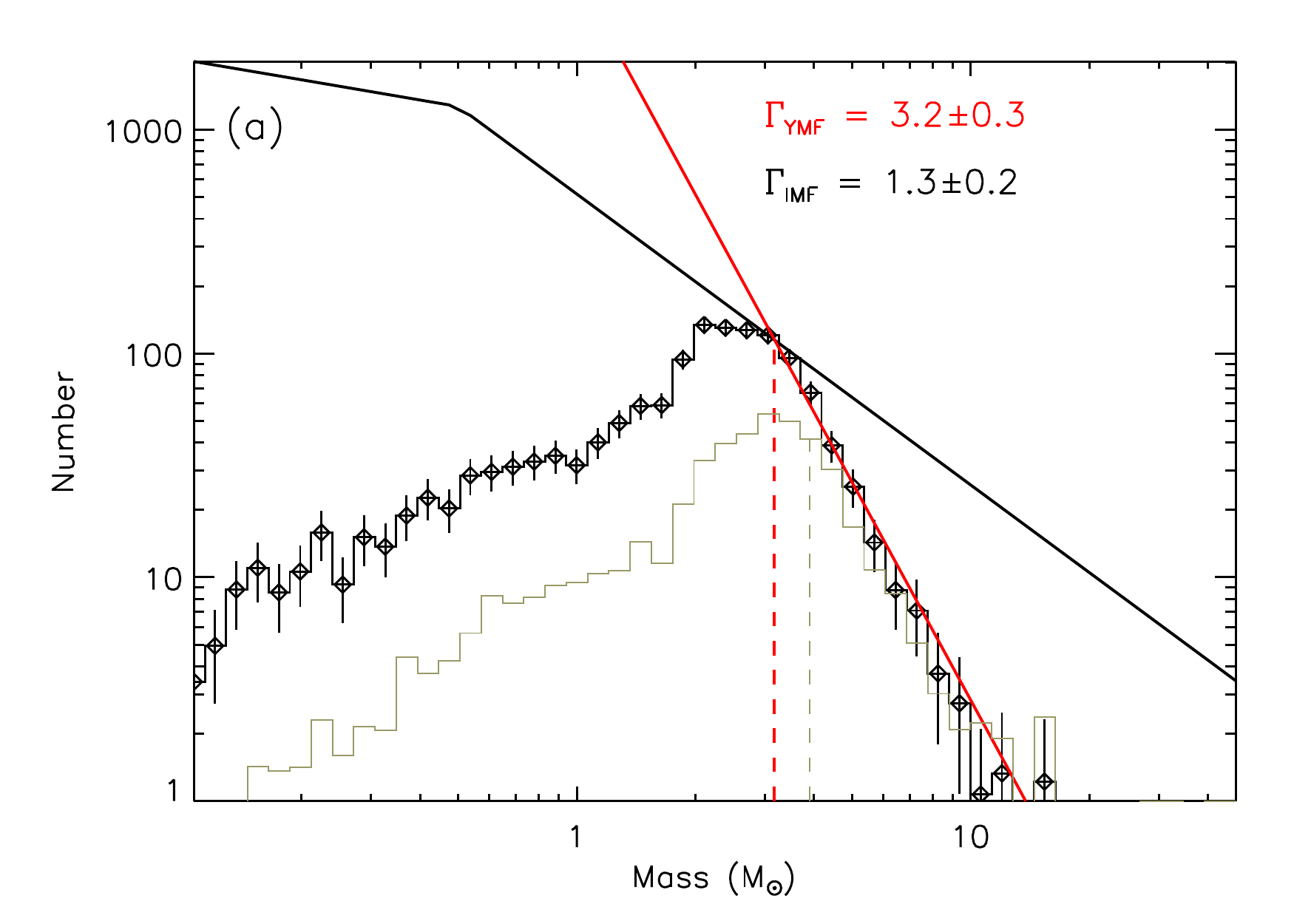}\\
\plotone{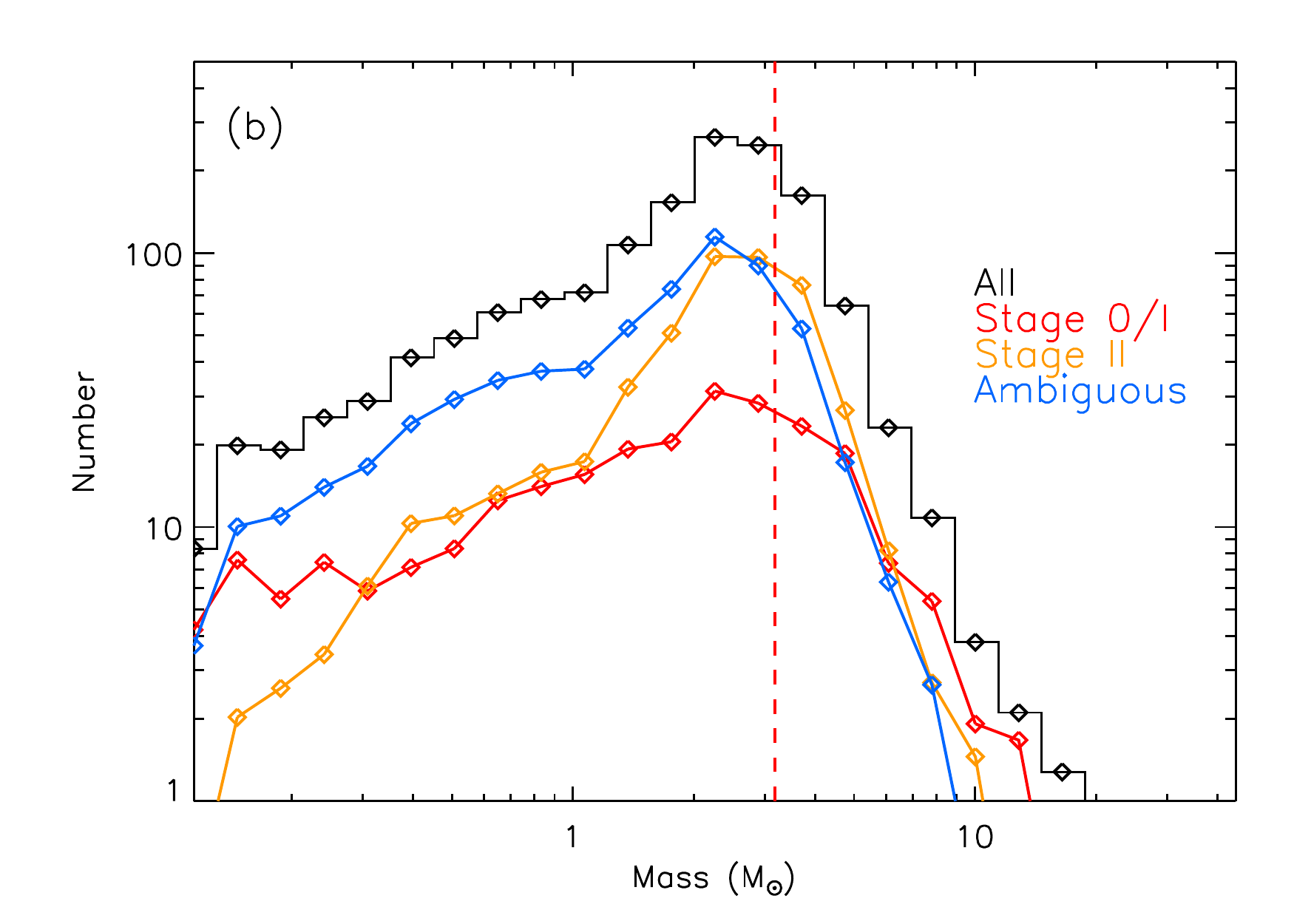}
\caption{({\it a}) YMF plot for all 1439 YSOs in the Catalog
  (histogram points with error bars). Two IMFs are 
  overplotted: power-law fit to the YMF
  (red) over the range $m\ge M_p$ and \citet{Kroupa} 
  IMF scaled to
  match the YMF at $M_p=3.1$~\Msun\ (dashed
  red line). The YMF from PW10 for 488 YSOs in the M17 SWex cloud
  is overplotted for comparison (gray histogram), and its critical
  mass $M_c=3.9$~\Msun\ (see text) is indicated (dashed gray line).
 ({\it b}) Carina YMF binned
  up by a factor of 2 and subdivided by evolutionary stage. 
\label{YMF}}
\end{figure}

Summing the individual probability distributions of $M_{\star}$ defined by the
set of models fit to each YSO, we construct the YSO mass function
\citep[YMF;][PW10]{SP07,P09,MK09} 
for all 1439 YSOs in the PCYC (Figure~\ref{YMF}). The YMF is
defined similarly to the stellar initial mass function (IMF),
$\Phi(\log{m})=dN/d\log{m}$, where $dN$ is the number of stars in the
(logarithmic) mass interval $(\log{m},\log{m}+d\log{m})$. The Carina
YMF exhibits a power-law form $\Psi(m)\propto m^{-\Gamma_{\rm YMF}}$
for $M_c\le m<10$~\Msun, departing from the power-law due to incompleteness
for $m < M_p\approx 3.1$~\Msun\ (Figure~\ref{YMF}a). The power-law
slope $\Gamma_{\rm YMF}=3.2\pm0.3$ is significantly steeper than the standard
Salpeter--Kroupa IMF slope $\Gamma_{\rm IMF}=1.3$ \citep{Kroupa}.
PW10 found a very similar YMF shape, with $\Gamma_{\rm YMF}=3.5\pm
0.6$, 
for a sample of 488 YSOs in M17 SWex, a GMC complex
extending ${\sim}50$ pc outward from the bright Galactic \hii region M17. 

%

In Figure~\ref{YMF}b, the YMF has been broken down by most probable
evolutionary stage. This shows that the overall
shape of the YMF is dominated by a narrow component from Stage
II objects, but also includes a broader component from Stage 0/I
objects. The YMF shape of the Ambiguous
sources is
a blend of these two components. (Stage III YSOs constitute only 6\% of
the PCYC and are not plotted.) Ambiguous YSOs dominate the 
(incomplete) low-mass range, which is expected because fainter sources are
more likely to lack 24 \um\ detections and other photometric
measurements of sufficient quality to tightly constrain their properties. Stage
0/I YSOs are preferentially detected over Stage II YSOs at lower masses because
the reddest objects are generally brighter in the mid-IR for a given
mass. The steep intermediate-mass slope of the YMF traces both Stage II
YSOs and Ambiguous YSOs that are most likely dominated by Stage II
objects in
this mass range, given the overall distributions of Stage 0/I versus
Stage II YSOs. This supports the idea that disk evolution is an
important selection effect that steepens the observed YMF slope (PW10).

SP10 considered the idea that rapid disk destruction in massive stars
could preferentially remove them from the YSO sample and affect the
observed mass function but concluded that this effect does not
significantly impact 
an IMF derived from the YSOs in Carina. This conclusion was based upon
a perceived similarity between the empirical $K_S$ and $[3.6]$ luminosity
functions of the Carina YSOs and the scaled $K$- and
$L$-band luminosity functions (KLFs and LLFs) for the Orion Nebula
Cluster \citep[ONC;][]{Muench}. However, 
\citet{MLL00} caution that LFs are not reliable tracers of IMFs
when the photometry is dominated by IR excess emission. To
illustrate the problem, consider the bright end of the KLF.
The few most luminous YSOs in the SP10 sample, which are also included in the
PCYC, have similar $M_K$ to $\theta^1$ Ori C, the most luminous star
in the ONC. But while the most massive YSOs in Carina are
${\sim}10$~\Msun\ objects, $\theta^1$ Ori C is an O star of ${\sim}40$~\Msun!
While the ONC is a young cluster, the majority of its stars have lost
their inner disks,
and the KLF used by \citet{Muench} to derive the cluster IMF was
dominated by emission from stellar photospheres. To better understand the IMF
traced by the YSOs in Carina, we rely instead on the RW06 models to
relate statistically the observed IR emission to stellar mass. Our
approach is subject to its own set of biases and systematics, but it has been
designed for the problem at hand and has been tested in numerous
other studies \citep[e.g.][]{BW08,P09,MK09}. We are also aided by the ability
to compare Carina to M17 SWex, where a similar sample of YSOs has been
analyzed using the same tools (PW10).

The YMF of 488 YSOs in M17 SWex from PW10 is overplotted in Figure~\ref{YMF}a
for comparison with the PCYC. The similarity between these YMFs is
striking;
the only obvious difference appears to be a lower mass
completeness limit for the PCYC.
Both Carina and M17
are 100-pc scale GMC complexes at similar
heliocentric distances, 
each containing several $10^5$~\Msun\ in molecular gas mass, and each
in the process of forming a large OB association. The sensitivity of
the GLIMPSE observations to YSOs in M17 SWex
was limited primarily by interstellar extinction
through the IR dark cloud. In contrast, the interstellar extinction to
the Carina YSOs is much lower, and the sensitivity is limited
instead by bright nebular emission produced by the \hii regions. 
It is not obvious, however, why high extinction should be a much
greater impediment to
YSO detections than bright nebulosity.
The implied difference
in sensitivity is dramatic, meaning that if
the YMF of PW10 is intrinsically similar to the PCYC, its completeness
at $m=3.1$~\Msun\ (corresponding to main-sequence B9 stars)
is as low as ${\sim}40\%$. 

The M17 SWex GMC
complex appears to be forming its first generation of stars and
lacks very massive, early O stars (PW10). The Carina molecular cloud complex has
been thoroughly disrupted by the process of forming tens of thousands
of stars, including 
some of the most massive stars known in the Galaxy
\citep{SB07}. 
The environmental differences between these two massive
star-formation regions lead us to expect that there are {\it
  intrinsic} differences between the YSO populations, and these
differences might be manifest in the YMFs.
In particular, we expect that the PCYC includes a larger
proportion of more evolved YSOs than the M17 SWex population. 

%
\begin{figure}
\plotone{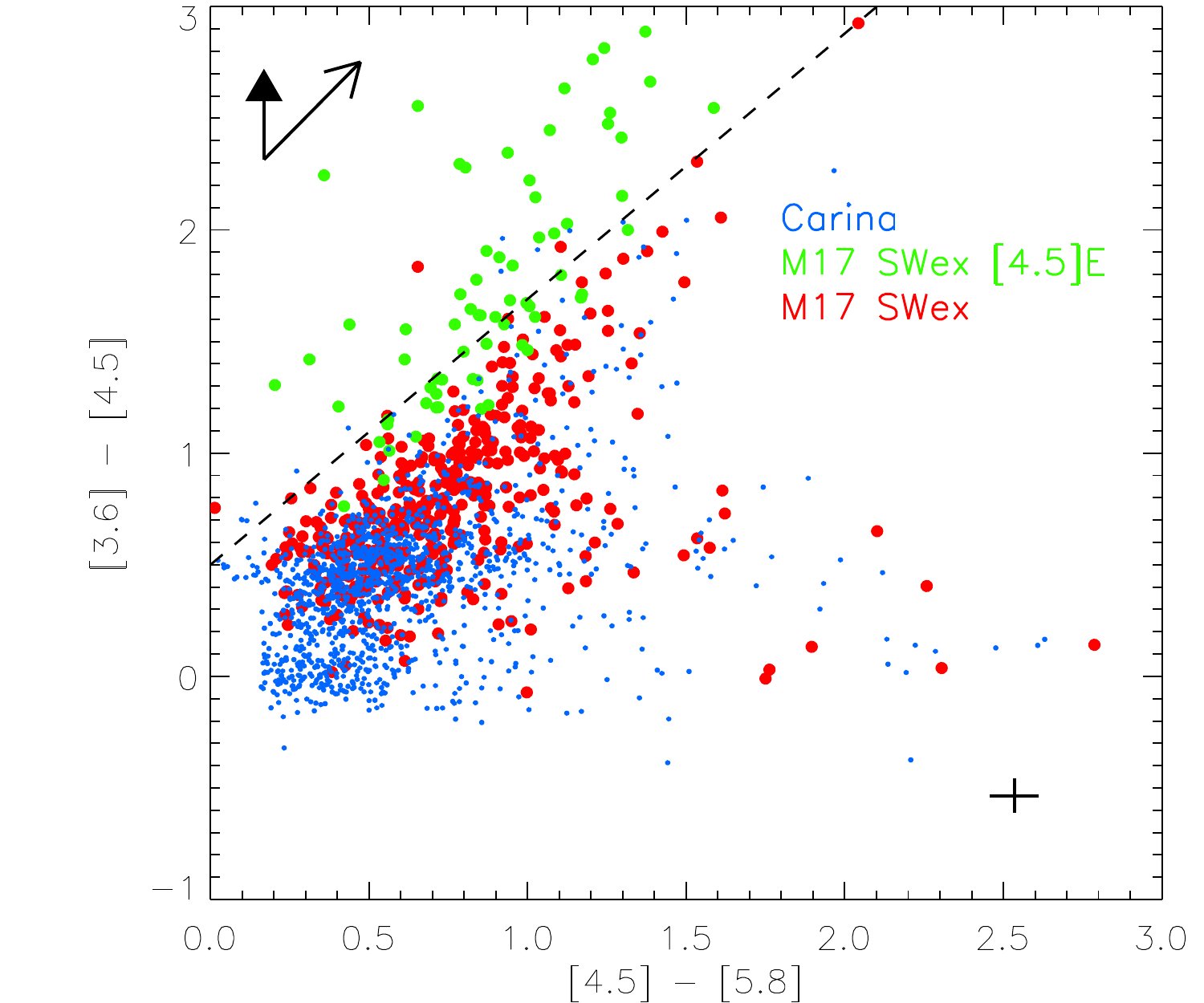}
\caption{IRAC color space enabling the separation of YSOs with excess
  4.5~\um\ ([4.5]E) emission. PCYC sources detected in these 3 bands are
  plotted along with sources from the M17 SWex YSO sample (PW10), with
  [4.5]E YSOs plotted in green. Candidate [4.5]E objects are located
  near or above the dashed line. 
  Reddening vectors
  for $A_V=30$~mag using the extinction laws 
  of \citet{WD01} and \citet{I05} are
  shown as open and filled arrows, respectively. The cross represents the
  typical photometric uncertainty.
\label{ugos}}
\end{figure}

A useful empirical comparison of the PCYC with the M17 SWex YSOs is
provided by the plot of the IRAC $[3.6]-[4.5]$ versus $[4.5]-[5.8]$
color space presented in Figure \ref{ugos}. PW10 identified a class of
predominantly Stage 0/I YSOs that exhibited excess 4.5~\um\ emission
(denoted [4.5]E)
with respect to the RW06 model SEDs and hypothesized that such sources
are unresolved analogs of the extended green objects (EGOs), candidate
molecular outflows from massive YSOs cataloged from the GLIMPSE images by
\citet{egos}. In this interpretation, such sources are expected to be
very young, with ages ${\la}10^5$ yr, since outflows are powered by rapid accretion \citep{RS00,SP07}. 
We find that
${\sim}50\%$ of [4.5]E candidates from PW10 are found 
above the line 
\begin{equation}
  [3.6] - [4.5] > 1.19([4.5]-[5.8])+0.5
\end{equation}
in Figure~\ref{ugos}; these ``strong'' [4.5]E sources represent 7\%
of the YSOs in M17 SWex, but ${<}1\%$ of the PCYC. SP10 searched the
\spitzer\ images 
of Carina for EGOs and found only four candidates. The
corresponding dearth of [4.5]E sources supports the conclusion that these
objects trace the same phenomenon as the EGOs. The difference between
the two regions could be
partly environmental; SP10 suggest  
that the harsh radiation fields permeating the Carina Nebula may
dissociate molecules in the outflows, removing the source of the EGO emission.
But this comparison
also indicates that very young YSOs with high accretion rates are less
frequent in Carina than in M17 SWex.
Indeed, only 17\% of the PCYC are unambiguous Stage
0/I YSOs, compared to 27\% Stage 0/I
YSOs in the M17 SWex sample (PW10).

The presence of a dusty, circumstellar disk (or an envelope that
will presumably evolve into a disk) is a
prerequisite for inclusion in 
the PCYC, hence the upper limit on the age of the YSOs is approximately the
disk lifetime. The ``canonical'' disk lifetime for a solar-mass,
classical T Tauri 
star is ${\sim}2$~Myr \citep{HLL01}. 
If the primary mechanism for destroying dust disks around intermediate-mass YSOs is
photo-evaporative dissipation driven by radiation from the central
star \citep{DH94,MMG02},
then the disk 
lifetime should decrease as a strong function of stellar mass. 
Disk lifetimes for intermediate-mass stars,
including Herbig Ae/Be stars, appear to be significantly shorter than
for T Tauri stars,
typically ${\la}1$~Myr \citep[][PW10]{JH07,P09}. 
The disk fraction among intermediate-mass stars may therefore be one
of the most sensitive chronometers available for massive
star-formation regions like Carina. 

PW10 showed that the shape of the
YMF depends critically on the mass-dependent disk fraction, and hence
the recent star formation history. Assuming 
a constant SFR, the observed YMF
slope is steepened by disk evolution above a critical mass
$m>M_c$ for which the disk lifetime $\tau_d(m)$ is less than the age
spread of the YSO 
population $\tau_0$. For $m\le M_c$, $\tau_d(m)\ge \tau_0$, the disk fraction
becomes unity, and the YMF traces the underlying IMF, presumably a
Salpeter--Kroupa slope ($\Gamma_{\rm 
  YMF} \rightarrow \Gamma_{\rm IMF}=1.3$). As a result,
$M_c$ decreases with time. 
Unfortunately, for the PCYC $M_c \le M_p=3.1$~\Msun, meaning
that the critical mass falls in the range where our observations are
incomplete. This limit means that the critical mass of the 
PCYC is lower than $M_c=3.9$~\Msun\ found by PW10 for the M17 SWex sample
(Figure~\ref{YMF}a), and this could be explained 
by disk evolution proceeding over a longer timescale in the Carina complex.
The PCYC does contain a significant population of Stage II/III YSOs with
relatively evolved disks.
This population is located near the line $[3.6]-[4.5]=0$ in Figure~\ref{ugos},
hence reddening by interstellar dust is negligible, and the lack of
IR excess emission at wavelengths ${\le}4.5$~\um\ suggests that the
inner disks have been cleared of dust.

Assuming that the underlying IMF of the Carina young stellar
population has a normal
Salpeter--Kroupa form, we can estimate the present-day SFR in Carina from
the PCYC. Scaling the IMF of the ONC \citep{Muench} 
to match the PCYC YMF where it departs from a power-law at
$M_p=3.1$~\Msun\  gives a prediction of  
${>}2\times 10^4$ YSOs to a limiting mass of 0.1~\Msun, corresponding to
a total mass in stars of ${>}1.6\times 10^4$~\Msun. These numbers are lower limits
due to incompleteness. Adopting a conservative upper limit of
$\tau_0\la 2$~Myr as the age spread 
sampled by the PCYC, we arrive at a {\it lower limit} on the recent SFR in
the Carina Nebula of ${>}0.008$~\Msun~yr$^{-1}$. If the Galactic SFR is a
few \Msun~yr$^{-1}$ \citep{MR10,RW10}, then the star 
formation activity of the entire Milky Way is equivalent to a few hundred
Carina complexes.

\section{Properties of the X-ray-Emitting YSO Population}\label{XYSOs}


The PCYC contains 410 X-ray-detected,
predominantly intermediate-mass YSOs, including 62 Stage 0/I YSOs
that are candidate X-ray-emitting protostars,
the largest such sample
compiled to date (Table~\ref{numbers}). 
Because Herbig Ae/Be stars 
fall into an X-ray ``desert'' between low-mass T Tauri stars with
convection-driven emission and OB stars with wind-driven emission, we
expect that many of the PCYC sources are 
intrinsically X-ray quiet. This motivates us to assume the null
hypothesis that X-ray emission apparently associated with
intermediate-mass YSOs is 
generally produced by an unresolved low-mass, T Tauri binary companion
\citep{BS06,NE11}.
In
such a scenario, the intermediate-mass primary is X-ray quiet but dominates the IR
emission, and the physical properties derived from the IR SED modeling
are expected to be uncorrelated with the presence of an X-ray-detected secondary.
This null hypothesis can therefore be disproved if we observe
statistically significant differences between the modeled properties of YSOs
with X-ray counterparts and those without.

The detection of a faint X-ray source is highly
dependent on location, because the sensitivity of the ACIS
observations varies significantly across 
the CCCP field \citep{CCCPcatalog}. 
The sensitivity limit of the PCYC has a
complicated positional dependence due to large spatial variations in
extinction and nebulosity.
Physical properties inferred from the IR
modeling can range over orders of magnitude, and observed X-ray
luminosity $L_X$ versus stellar mass (or $L_X$ versus $L_{\rm
  bol}$) relations have over 
an order of magnitude scatter \citep[e.g.][]{TP05}. 
Potential correlations between properties derived independently from
X-ray and IR observations are likely to be overwhelmed by severe noise. 
To avoid this problem, we divide the PCYC into CCCP-matched
(PCYC-X) and unmatched (PCYC-U) subsamples and compare  physical
properties derived from SED fitting between these subsamples.
The PCYC-U includes both
YSOs that are intrinsically X-ray quiet and YSOs that happened to fall below the
CCCP detection limit. 
There is likely to be substantial overlap in
the physical parameters between these subsamples. Rather than attempt
the potentially intractable task of correcting for the incredibly
complicated, competing
sensitivity variations between the X-ray and IR sample selections, we regard
sample overlap as an inevitable, 
additional source of noise in our comparisons. At best, we might hope
to discern some  
general, qualitative trends from the ensemble population, given the large sample
size. 

%
\begin{figure}
\plotone{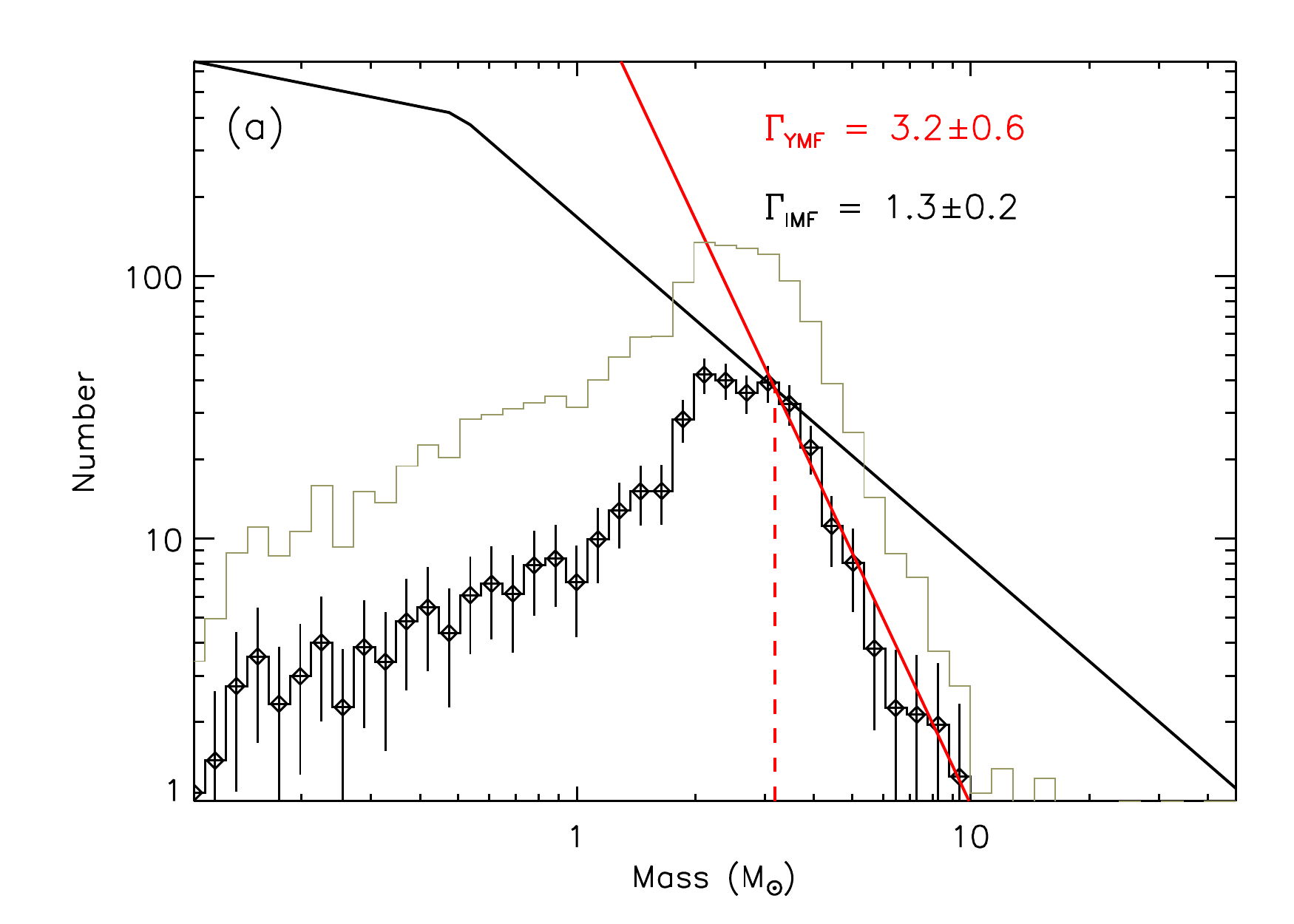}\\
\plotone{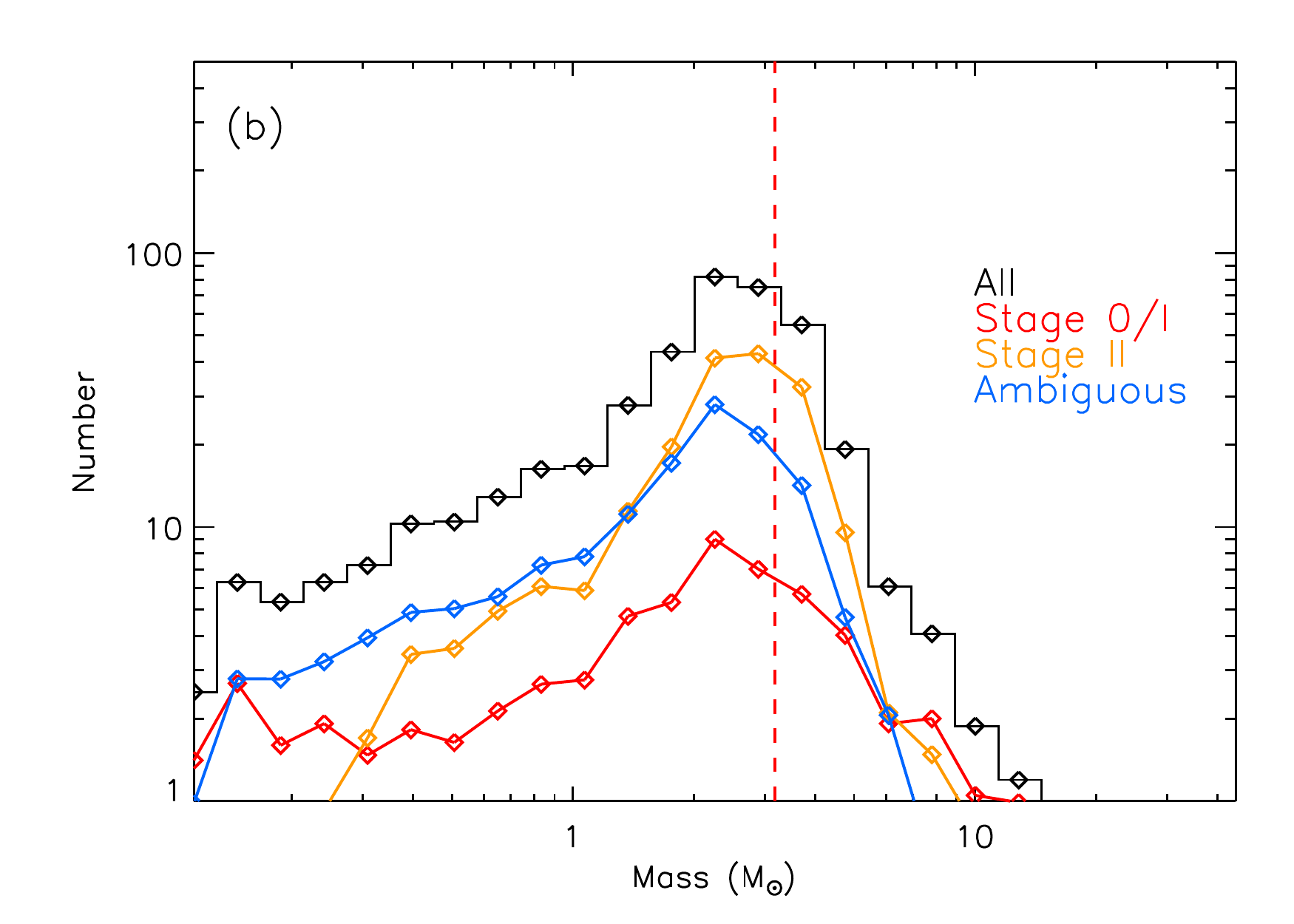}
\caption{Same as Figure~\ref{YMF}, except only the 410 YSOs with X-ray
  counterparts are included in the X-YMF. The gray histogram shows the
  full PCYC YMF from Figure~\ref{YMF}.
\label{YMF_ACIS}}
\end{figure}

The YMF of the 410 YSOs in the PCYC-X (the X-YMF)
is plotted in Figure~\ref{YMF_ACIS}. 
The intermediate-mass slope $\Gamma_{\rm
  YMF}=3.2\pm0.6$ and power-law departure mass $M_p=3.1$~\Msun\ 
(Figure~\ref{YMF_ACIS}a) are consistent with 
the full PCYC YMF. 
This similarity is striking, given that we might have expected the X-YMF
to skew toward lower-mass YSOs drawn from the T Tauri population. 
Instead, there appears to be a subtle trend in the {\it opposite}
direction, with the $m<2$~\Msun\ mass range suppressed in the
X-YMF. 
This trend is also apparent when the X-YMF is plotted
by evolutionary stage (Figure~\ref{YMF_ACIS}b). 


%
\begin{figure}
\epsscale{0.7}
\plotone{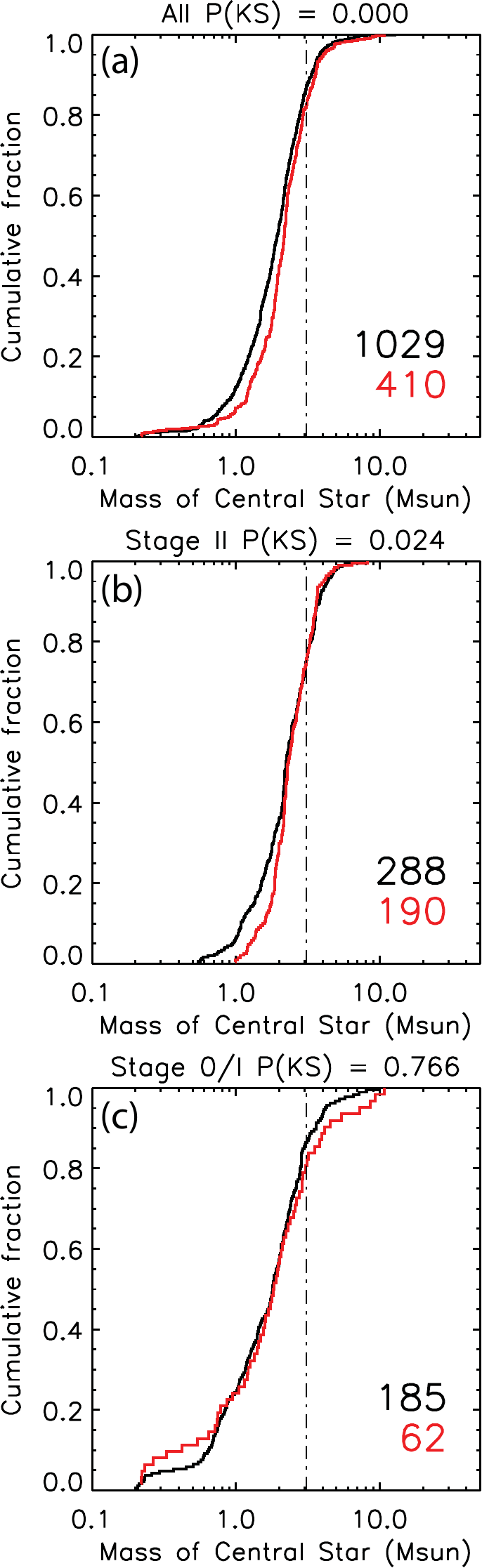}
\caption{Cumulative mass functions comparing X-ray detected (red) and
  undetected (black) YSOs: (a) All YSOs in the PCYC, (b) Stage II YSOs, and 
  (c) Stage 0/I YSOs. 
  The number of YSOs in each curve is
  printed in each panel, and each panel header gives the probability
  from a KS test that the two samples are drawn from the same
  underlying distribution. The vertical dash-dotted lines indicate the
  YMF power-law departure mass, $M_p=3.1$~\Msun\ (see Figure~\ref{YMF}). 
\label{cum_MF}}
\end{figure}

Differences between the PCYC-X and PCYC-U are easier to discern
from the cumulative mass distributions (Figure
\ref{cum_MF}), constructed from the characteristic $\langle
M_{\star}\rangle$ for each YSO (column 4 of Table \ref{mod}).
Two curves are plotted in each panel, representing the PCYC-U and the
PCYC-X (red) subsamples. The panel headers give the probability $P({\rm KS})$ 
 from a
Kolmogorov--Smirnov (KS) test
that the two curves are drawn from the same underlying
distribution.  
Figure~\ref{cum_MF}{\it a} shows that the PCYC-X as a whole
includes a significantly lower fraction of sources with
$m<2$~\Msun. Figure~\ref{cum_MF}{\it b} reveals that this difference is produced by
the Stage II sources. Evidently the Vela--Carina \spitzer\ observations are
 more sensitive than the CCCP observations to low-mass, disk-bearing stars.
A KS test shows no significant difference between the
cumulative mass distributions of the two samples of Stage 0/I sources,
although the two curves do appear to diverge for $m>M_p$ (Figure
\ref{cum_MF}{\it c}).

\subsection{Intermediate-Mass, Disk-Dominated YSOs with X-ray Counterparts}

\begin{figure*}
\plotone{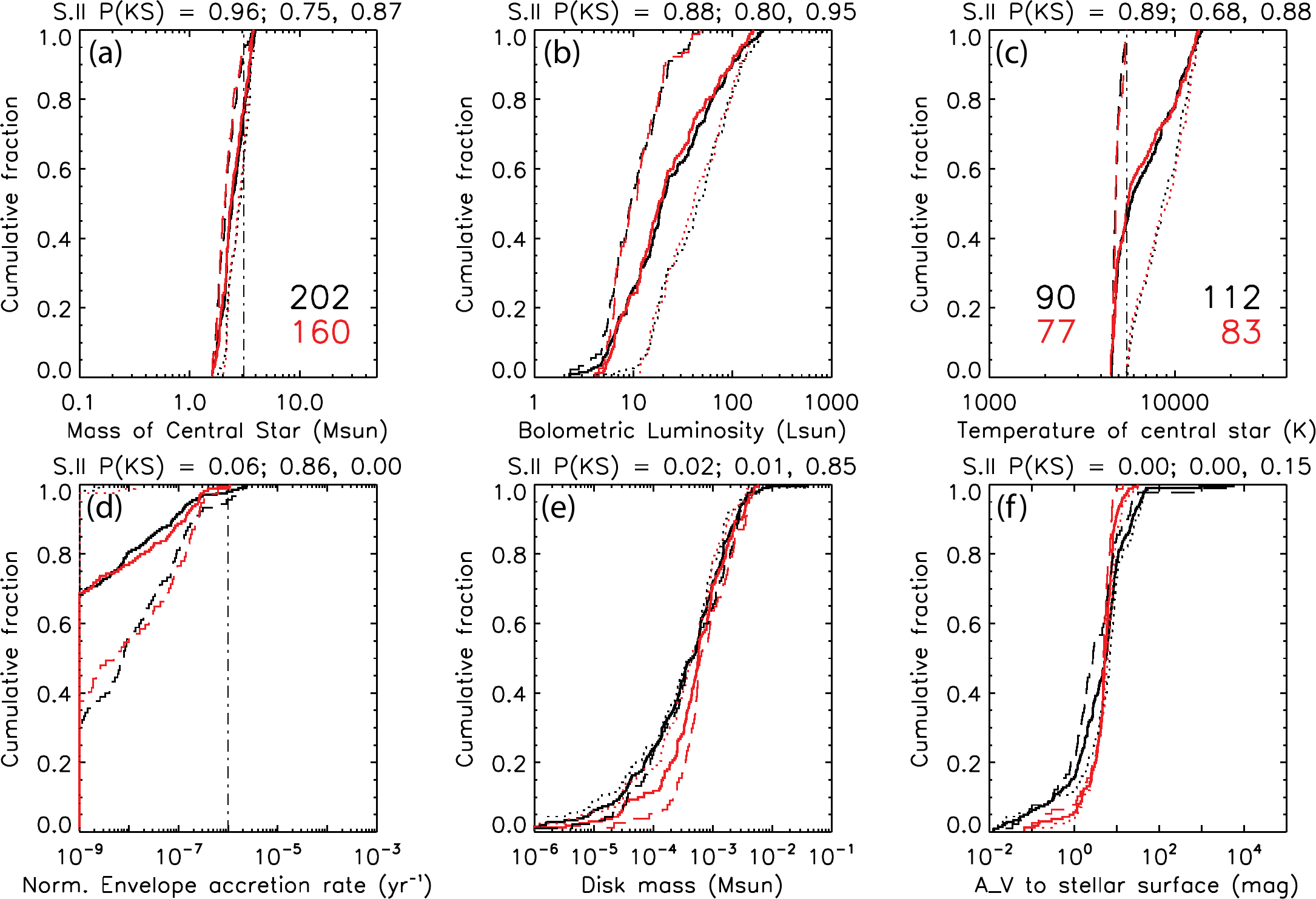}
\caption{Cumulative distribution functions of 6 key YSO model parameters (see
  discussion in text) for Stage II YSOs with $1.6~{\rm M_{\sun}}\le \langle
  M_{\star}\rangle<4$~\Msun. The dividing line at $\dot{M}_{\rm
    env}/M_{\star}=10^{-6}$~yr$^{-1}$ between Stage II and
  Stage 0/I YSOs is the vertical dash-dotted line in
  panel {\it d}. Colors and annotations are the same as in 
  Figure~\ref{cum_MF}. The distribution functions are further subdivided
  by stellar temperature; long-dashed curves for $\langle
  T_{\star}\rangle<5500$~K and dotted curves for $\langle
  T_{\star}\rangle \ge 5500$~K (the fiducial temperature is the vertical
  dash-dotted line in panel {\it c}). Three KS probabilities
  are given in each panel header: full distribution; low-$T_{\star}$,
  high-$T_{\star}$ sub-distributions.
\label{SII_comp}
}
\end{figure*}

In Figure~\ref{cum_MF}{\it b} the mass distributions
of both the PCYC-X and PCYC-U samples appear to be very
similar for $m \ga 2$~\Msun. In Figure~\ref{SII_comp} we present a
comparison of 6 key YSO model parameters for the unambiguous Stage
II YSOs in the intermediate-mass range $1.6~{\rm M_{\sun}} \le\langle
M_{\star}\rangle < 4$~\Msun, equivalent to A through late B
stars on the main sequence. The parameters plotted are stellar mass
$\langle M_{\star}\rangle$, bolometric luminosity $\langle L_{\rm
  bol}\rangle$ (which includes luminosity derived from accretion), stellar
temperature $\langle 
T_{\star}\rangle$, normalized
envelope accretion rate $\langle \dot{M}_{\rm env}/M_{\star}\rangle$,
disk mass $\langle M_{\rm disk}\rangle$, and total extinction to the
stellar surface $A_{V,t}$ (defined as the sum of the contributions to extinction
produced by dust along the line of sight $A_V$ and circumstellar dust
$A_{V,c}$). The characteristic values $\langle X\rangle$ of the stellar parameters
$X=(M_{\star}$, $L_{\rm bol}$, $T_{\star}$) were calculated using
Equation~(\ref{stellar}), and the characteristic values of the
circumstellar/environmental parameters $\langle Y\rangle$ were
calculated using Equation~(\ref{circumstellar}).

%

We further subdivide our intermediate-mass Stage II sample by
stellar temperature, into sets with $\langle T_{\star}\rangle<5500$~K (167
YSOs, 77 in the PCYC-X) and $\langle T_{\star}\rangle \ge 5500$~K (195 YSOs,
83 in the PCYC-X). We were motivated to make this division by the hypothesis that
intermediate-mass stars with cooler photospheres are more
likely to have convective atmospheres and perhaps generate X-rays by a
mechanism analogous to that of lower-mass T Tauri stars \citep[e.g.][]{KS04,TP05}. 


We find that none of
the mass, luminosity, or temperature distributions
(Figures~\ref{SII_comp}{\it abc}) shows any significant
correlations with the 
presence of X-ray emission. 
The normalized envelope
accretion rates are low by definition for Stage II YSOs, and the
distributions shown in Figure~\ref{SII_comp} confirm that these
objects lack significant infalling envelopes. We do observe that the
distribution functions for the low-$T_{\star}$ division (dashed
curves) have
systematically lower luminosity and 
mass compared to the high-$T_{\star}$ division, which is expected from
PMS stellar evolutionary tracks. 

There are no statistically significant differences in the cumulative
distribution functions of the model parameters among the
high-$T_{\star}$ division that correlate with X-ray
emission.\footnote{The one exception is normalized envelope accretion
  rate, $P({\rm KS})=0.00$ in Figure~\ref{SII_comp}{\it d}; however this is
  not a meaningful result 
  because the envelope accretion rates are near zero in Stage II YSOs,
  and this parameter is not expected to be well constrained by the SED
  fitting.}
This result is consistent with our null 
hypothesis; we cannot rule out the possibility that X-ray emission
associated with Stage II YSOs with $\langle 
T_{\star}\rangle\ge 5500$~K actually originates in unresolved, lower-mass T Tauri
companions. 
In contrast, the distribution functions of $M_{\rm disk}$
and $A_{V,t}$ (Figures 7{\it ef}) definitively fail the KS test for
the low-$T_{\star}$ division and 
demonstrate that X-ray  
emission is {\it not} a completely 
random phenomenon among the intermediate-mass Stage II YSOs. 
PCYC-X sources are significantly {\it underrepresented} among YSOs with the
lowest disk masses, $\langle M_{\rm
  disk}\rangle<10^{-3}$~\Msun\ (Figure~\ref{SII_comp}{\it e}). 
We have uncovered a population of intermediate-mass PMS stars with
massive disks and {\it intrinsic X-ray emission}, but the X-ray
emission apparently
dies off as the disks evolve.
This result seems diametrically opposed to the
observational paradigm in which X-ray luminosity from low-mass PMS stars
increases as these objects evolve from classical T Tauri stars to
weak-lined T Tauri stars  \citep{TP05,XEST}, a transition that reflects
the cessation of disk-fueled accretion that may suppress X-ray
emission \citep{KS04}.

The main result illustrated by Figure~\ref{SII_comp} is that disk mass
correlates far more strongly 
with X-ray emission than any of the stellar parameters,
demonstrating that {\it rapid disk evolution provides a
highly sensitive chronometer for intermediate-mass PMS stars.}
The apparent paradox with T Tauri X-ray emission is resolved when we
recall that this trend is 
observed {\it only} among our carefully selected sub-division of
intermediate-mass PMS stars with $\langle M_{\star}\rangle \ge 1.6$~\Msun\ and $\langle
T_{\star}\rangle<5,500$~K. These stars will 
presumably evolve into Herbig 
Ae/Be stars, at which point their X-ray brightness would fall below the
CCCP detection limits \citep{BS06,CCCPcatalog}. 
At present, they are very young objects with relatively
massive disks, still contracting toward the main sequence, and they have
spectral types of G2 or later. They could thus be regarded as more
massive, more luminous analogs of classical T Tauri 
stars.

A direct imprint of circumstellar disks on X-ray emission can be
inferred from the 
distribution of total extinction ($A_{V,t}$; Figure~\ref{SII_comp}{\it f}).
The top 5\% of the PCYC-U reach absorptions up to ${\sim}100$ times
higher than the top 5\% of the PCYC-X. These sources are probably
high-inclination systems, where {\it nearly
edge-on disks completely absorb any X-ray emission along the sightline
to the central stars.}

\subsection{Candidate X-ray-Emitting Protostars}

\begin{figure*}
\plotone{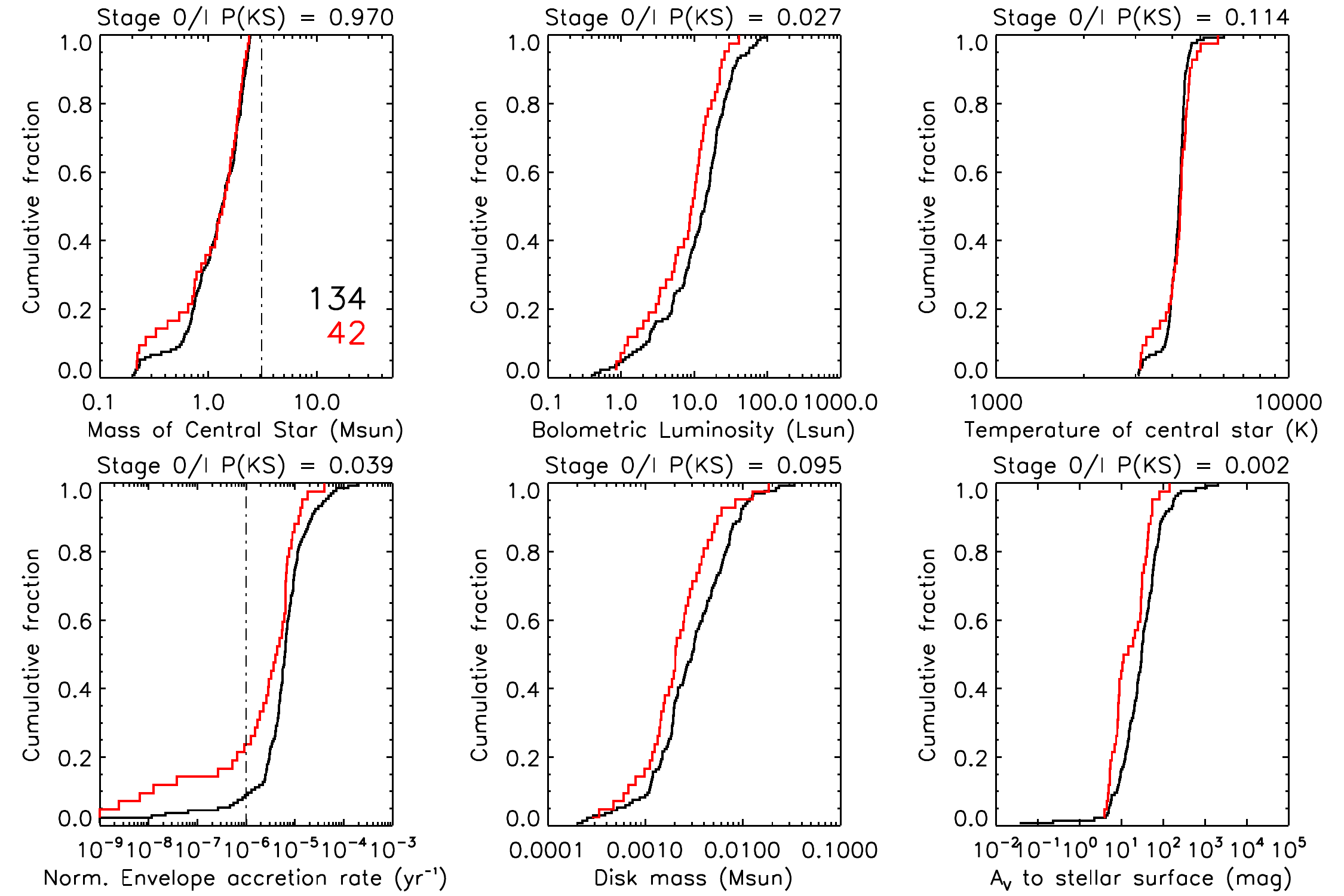}
\caption{Cumulative distributions of 6 key YSO model parameters for
  Stage 0/I YSOs with $\langle M_{\star}\rangle<2.5$~\Msun. Colors,
  annotations, and fiducial lines are the same as in
  Figs.~\ref{cum_MF} and \ref{SII_comp}.
\label{S0I_comp}}
\end{figure*}

The 62 unambiguous Stage 0/I YSOs in the PCYC-X subsample are
presented in Table~\ref{protostars}, which lists values for some of
the YSO model parameters along with 3
observed X-ray quantities; total-band (0.5--8~keV) net counts,
median energy ($E_{\rm median}$), and energy flux ($F_t$).
As Figure~\ref{cum_MF}{\it c} shows, the mass
distributions of Stage 0/I YSOs are very similar between the PCYC-X
and PCYC-U subsamples for $\langle M_{\star}\rangle<M_p$. To investigate
which properties other than mass may correlate with X-ray emission from
protostars, in Figure~\ref{S0I_comp} we present distribution functions
comparing the same 6 YSO model parameters 
used to analyze the intermediate-mass Stage II population
(Figure~\ref{SII_comp}), this time
plotting the unambiguous Stage 0/I
YSOs with $\langle M_{\star}\rangle<2.5$~\Msun.

%

KS tests reveal that the mass distributions for these low-mass Stage
0/I YSOs are 
statistically the same between the PCYC-X and PCYC-U
(Figure~\ref{S0I_comp}{\it a}), yet the
luminosity distributions are 
significantly different, with the X-ray emitters systematically
preferring {\it lower}
values of $L_{\rm bol}$ (Figure~\ref{S0I_comp}{\it b}). Hence we may
deduce that {\it X-ray 
  detected Stage 0/I YSOs are less luminous for a given stellar mass.}
Furthermore,
both the normalized envelope accretion rates $\dot{M}_{\rm
  env}/M_{\star}$
and total absorption 
$A_{V,t}$ (columns 6 and 7 of Table~\ref{protostars}) show
parameter distributions skewed to significantly {\it lower} values
for the X-ray emitters, while 
$T_{\star}$ is marginally higher for the X-ray emitters.
As protostars
contract, they become hotter and less luminous, their accretion rates
slow, and outflow-carved, bipolar cavities grow in their envelopes,
reducing the obscuration to the central star. 
Taken together, the 
distributions shown in Figure~\ref{S0I_comp} demonstrate that {\it
 brighter X-ray emission correlates with more evolved Stage 0/I YSOs.} 

Previous identifications of candidate X-ray emitting Class 0/I YSOs in
nearby star formation regions have observed
characteristically hard 
X-ray spectra, with $E_{\rm median}\ga 2.5$~keV
\citep{KG07,COUPprotostars}. 
X-ray spectra observed from embedded protostars 
are expected to be hard because soft X-rays are
preferentially absorbed both by the dense circumstellar envelopes and
the surrounding dense molecular cloud cores \citep{KG07}.
About half (26/62) of our sample of candidate X-ray emitting Stage 0/I
YSOs have hard X-ray counterparts in the CCCP catalog ($E_{\rm
  median}>2$~keV); these objects are high-probability candidate X-ray
emitting protostars, grouped together in Table~\ref{protostars}. 
The remaining Stage 0/I YSOs in Table~\ref{protostars} are matched to
soft X-ray sources, more consistent with expectations for Class II/III objects
\citep{TP05,COUPprotostars}. 
YSOs in this latter group may be in transition between the envelope-dominated and
disk-dominated phases. Many of them would not have been selected as
Class 0/I objects by the conservative mid-IR empirical criteria of
\citet{COUPprotostars}, for example. 
The model-based Stage 0/I
classification, however, indicates that the YSOs in
Table~\ref{protostars} are still very young (they possess envelopes);
whether or not they 
are identified as ``candidate protostars'' is primarily a semantic choice. 
As expected, the IR SED modeling indicates that the soft X-ray group is less
embedded than the hard X-ray group, with 
median $A_{V,t}= 10$~mag compared to 30~mag (column 7 of Table~\ref{protostars}).
Recalling that most YSOs in the Carina Nebula are found outside of
dense molecular clouds (\S\ref{spatial}), the typical absorption may
also be less than expected from observations of YSOs in other
star-forming molecular 
clouds that have not been disrupted by feedback from extremely massive stars.

\section{Butterfly Collection: Intriguing YSO Sub-Clusters}\label{butterflies}

In this section, we identify and briefly discuss YSOs found in
candidate embedded sub-clusters in the CCCP field.
These YSOs are highlighted, with their group membership identified, in
Tables~\ref{obs} and \ref{mod}. 
There are a few caveats to remember. The
properties of individual 
YSOs may not be well-constrained by the RW06 models. Sources might be
multiple or otherwise confused in our multiwavelength datasets, and
apparent spatial matches between CCCP sources and Vela--Carina sources do not
guarantee physical association. We
believe that these clusters are
sufficiently interesting individual
``butterflies'' among the larger stellar population in the Carina
complex to merit targeted follow-up multiwavelength 
observations that would provide higher spatial resolution, wider wavelength
coverage, and/or spectroscopy.

\subsection{A Perfectly Ordinary Stage 0/I YSO Associated with an
  Extraordinary Obscured X-ray Cluster}

%
\begin{figure}
\plotone{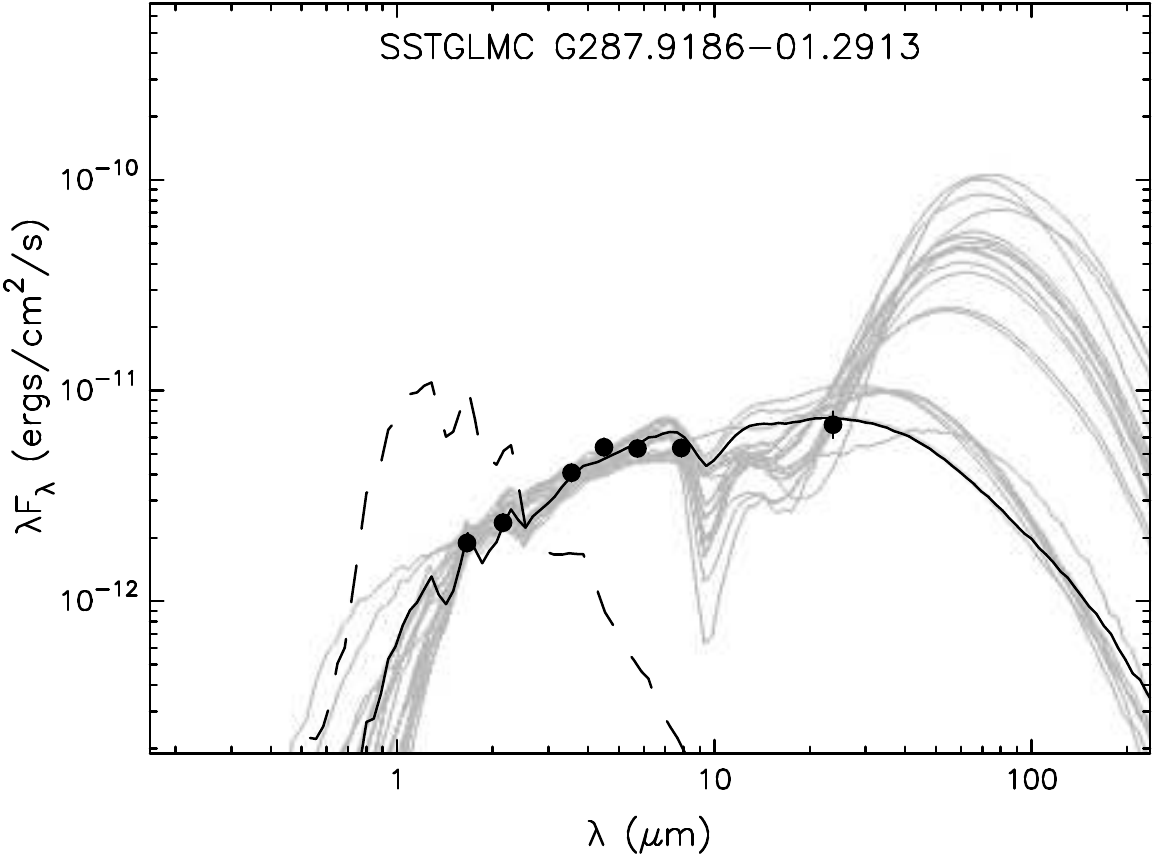}\\
\plotone{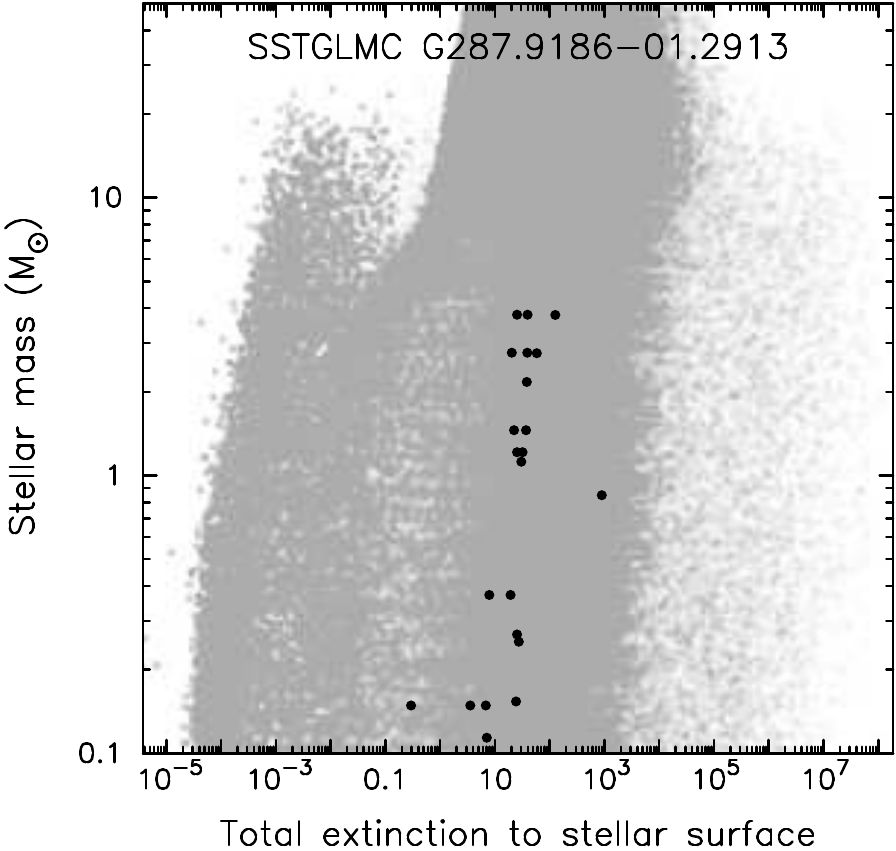}
\caption{{\it Top:} SED of PCYC 699 (points) and well-fit
  YSO models (solid curves). The heavy curve represents the best-fit
  model, and the dashed curve is the stellar photosphere of this model
  as it would appear in the absence of circumstellar extinction
  (removing $A_{V,c}$). {\it Bottom:} Plot of parameters $M_{\star}$
    versus $A_{V,c}$ for the set of well-fit YSO models. The
      gray shaded area represents the full parameter space sampled by
      the RW06 YSO models.
\label{Y699}}
\end{figure}
\citet{CCCPintro} report the discovery of a group of 4 X-ray sources
in the South Pillars region that are remarkably luminous, hard, and
highly-obscured. Three of these 
sources were matched to a single Vela--Carina Catalog source,
G287.9186-01.2913 \citep{CCCPcatalog}, or PCYC source 699
(Tables~\ref{obs}--\ref{protostars}). The primary CCCP match to PCYC
699 is CXOGNC J104451.91-602511.9, the brightest ($F_t=5\times
10^{-13}$~erg~cm$^{-2}$~s$^{-1}$) and
second-hardest ($E_{\rm median}=5.2$~keV) X-ray source associated with any
candidate protostar in the Carina Nebula
(Table~\ref{protostars}). This source is  
variable and spectacularly luminous in X-rays, with
$L_X=0.5$~\Lsun, corrected for $A_{V,t}=250$~mag total absorption
\citep{CCCPintro}. In stark contrast, there is nothing remarkable
about PCYC 699, save perhaps for the fact that its physical parameters are
more poorly constrained than usual. The RW06 model fits to the SED of PCYC
699 are plotted in Figure~\ref{Y699} along with a parameter plot
showing $M_{\star}$ versus $A_{V,c}$ for each fit. A degeneracy in the
long-wavelength SED beyond 24~\um, where we have no data,
produces 2 distinct families of model fits (Figure~\ref{Y699}):
intermediate-mass YSOs with relatively high circumstellar extinction
and low-mass 
YSOs with lower circumstellar extinction.
The model fits corresponding to the highest extinction are consistent with the
absorption of the X-ray source, but even taking into account the wide range
of stellar mass and luminosity allowed by the model fits, there is no
interpretation that would allow PCYC 699 to be considered a massive
YSO. It is a relatively faint 24~\um\ source, and, unlike the most
luminous YSOs in the PCYC, it is neither an \msx\ source nor an IRAS
source. Thus we are left 
with a mystery surrounding the origin of the bright X-ray
emission. The X-ray properties indicate a luminous, obscured, compact massive star
cluster, but there is no commensurate IR emission.

\subsection{Five Prominent YSO Sub-Clusters or Groups}

%
\begin{figure*}
\plotone{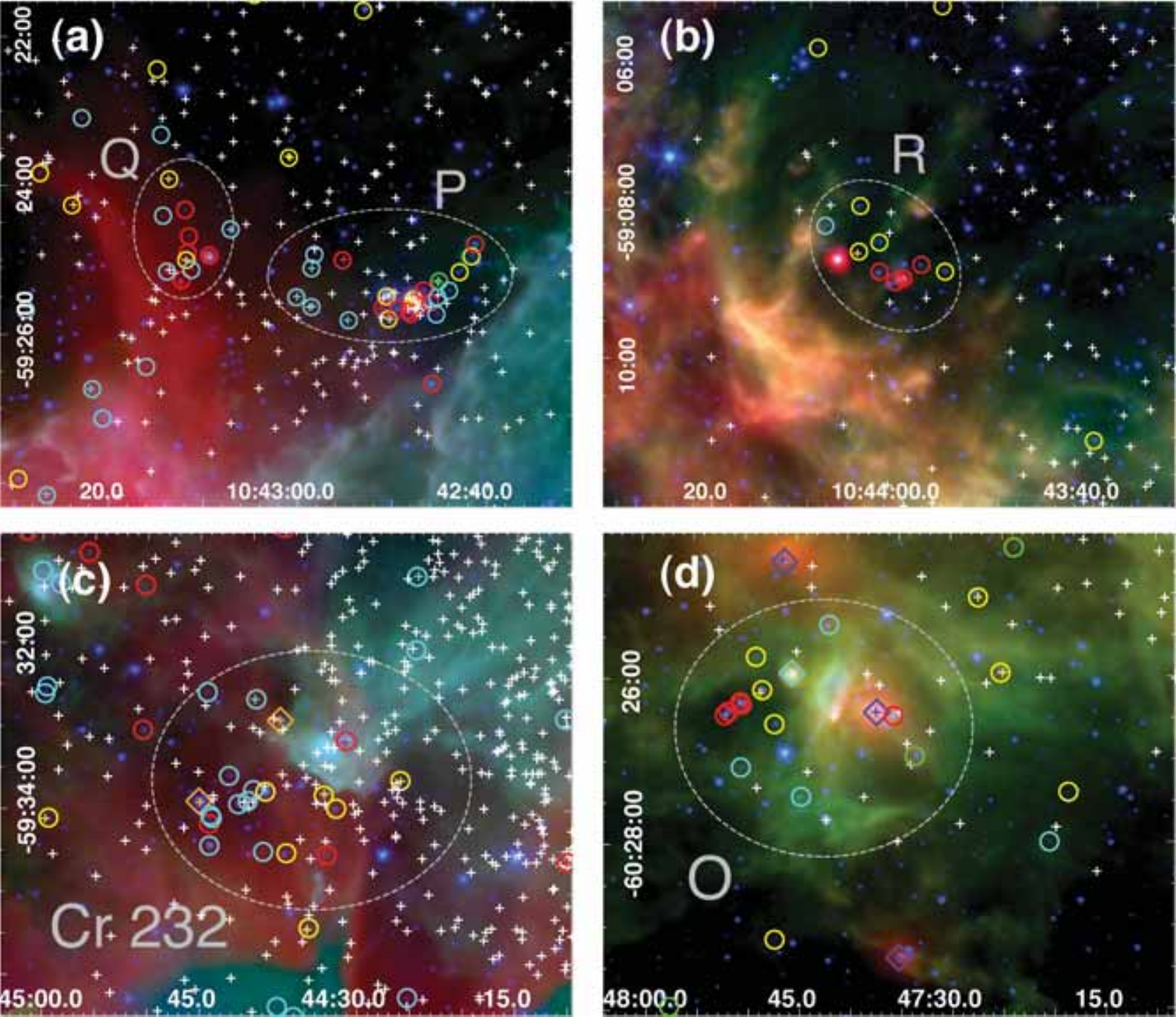}
\caption{\spitzer\ color images (red = 24~\um, green = 8.0~\um, blue =
  4.5~\um) highlighting the 5 tight groupings or
  clusters of YSOs discussed in the text. In all panels, YSOs are
  marked by circles color-coded by most probable evolutionary stage as
  in Fig.~\ref{image}, and crosses mark the positions of CCCP X-ray
  catalog sources. Where an X-ray source has been matched to a YSO,
  the crosses have been colored to match the YSO
  classification. Dashed ovals have been drawn subjectively to enclose
  the YSOs that appear associated with each grouping. Coordinate axes
  are equatorial (J2000). In
  panel (c), two cataloged, X-ray-emitting early O stars in Cr 232 are
  marked by
  orange diamonds and crosses; these are HD 93250 (left) and HD 303311
  (right). Regions at the bottom edge of this image appear turquoise
  due to saturation of the 24~\um\ nebular emission.
  In panel (d), three candidate X-ray detected OB stars from
  P11 are marked by blue diamonds and crosses, and a candidate 
  ultracompact \hii region is highlighted with a cyan diamond.
\label{gallery}}
\end{figure*}
In Figure~\ref{gallery} we present \spitzer\ multicolor images
zoomed in on 5 prominent YSO clusters, 3 of them new
discoveries from the combined \spitzer/Vela--Carina and CCCP observations. Extending the
nomenclature of the \spitzer\ clusters 
identified by SP10, we call the new clusters Spitzer P, Q, and R
(Table~\ref{mod}).

{\it Spitzer P and Q:} $(\alpha_{2000},\delta_{\rm
  2000})=(10^h43^m,-59\degree 26\arcmin)$. These two clusters appear
in projection to be close neighbors (Figure~\ref{gallery}{\it a}),
located ${\sim}8\arcmin$ 
northwest of Tr 14. Both contain a high fraction of unambiguous Stage
0/I YSOs and show no sign of dynamical relaxation, suggesting extreme
youth. The location of Spitzer P corresponds to a C$^{18}$O core
(Figure~\ref{image}), and its dense center contains a concentration of
X-ray sources (Group 4; F11). 
Spitzer Q was not
designated an X-ray group by F11, although it also
coincides with an enhanced density of X-ray sources
(Figure~\ref{newimage}). Possibly the most massive X-ray-emitting Stage
0/I YSO in the CCCP field is found in Spitzer Q. PCYC 179 is an
extremely red {\it Spitzer} source, detected by
\msx\ but not by 2MASS (Table~\ref{obs}), with $\langle L_{\rm bol}\rangle\sim
4\times 10^3$~\Lsun\ and a very high envelope accretion rate, $\langle
\dot{M}_{\rm env}\rangle\sim 
2.5\times 10^{-3}$~\Msun~yr$^{-1}$ (Table~\ref{mod}). The associated
CCCP source is very hard, $E_{\rm median} = 4.0$~keV
(Table~\ref{protostars}). If the intrinsic X-ray
emission is the soft spectrum produced by a normal massive stellar wind
\citep{LW80}, then it must be very luminous, with the soft counts almost completely
absorbed. Alternatively, a more 
exotic emission mechanism may produce in intrinsically hard X-ray
spectrum, or the CCCP source is an
unresolved, low-mass companion.

{\it Spitzer R:} $(\alpha_{2000},\delta_{\rm
  2000})=(10^h44^m,-59\degree 08\arcmin)$. In the extreme northern
corner of the CCCP field we have discovered another very young, tight
grouping of YSOs called Spitzer R (Figure~\ref{gallery}{\it b}). This
cluster is roughly centered 
on a CO peak that was {\it not} detected as a C$^{18}$O core by
\citet[][and see Figure~\ref{image}]{YY05}, and it appears to be forming
intermediate-mass stars but not massive stars. There is no associated
X-ray cluster, but this may simply be a sensitivity effect given that
Spitzer R lies close to the edge of the CCCP field and was observed
far off-axis by ACIS, where the sensitivity is reduced \citep{CCCPcatalog}.

{\it Cr 232:} $(\alpha_{2000},\delta_{\rm
  2000})=(10^h44.5^m,-59\degree 34\arcmin)$. Cr 232 is a well-known
young massive cluster to the west of Tr 14, containing an O3.5 V((f))
star, HD 93250, and an O5 V star, HD 303311 
\citep[][and references therein]{NS06a}. Cr 232 is a large X-ray cluster
(F11) apparently centered near the end of a broad dust
pillar oriented toward $\eta$~Car and Tr 16 (Figure~\ref{gallery}{\it
  c}). Cr 232 is undoubtedly richer than the newly-discovered {\it Spitzer}
YSO clusters. Its location suggests that its formation may have been
triggered recently by feedback from the massive stars in Tr 16.
Apparently in the middle of the dust pillar associated
with Cr 232 we detect a candidate X-ray-emitting, highly-embedded,
massive Stage 0/I YSO, 
PCYC 556 (Table~\ref{mod}), similar to PCYC 179 in Spitzer Q, but the
associated X-ray source is fainter by 0.4 dex and has $E_{\rm
  median}=1.9$~keV, placing it among the soft X-ray group in
Table~\ref{protostars}. PCYC 556 is among the reddest IRAC sources in
the PCYC ($[3.6]-[4.5]=1.4$~mag, $[5.8]-[8.0]=1.0$~mag;
Table~\ref{obs}). 
This relatively soft X-ray emission
apparently originating from an embedded massive star suggests a highly
luminous, intrinsically soft X-ray spectrum, perhaps powered by shocks
in a strong stellar wind or accretion flow.
Alternatively, the CCCP source may 
actually be a less obscured star that happens to be confused with the
bright IRAC source.

{\it Spitzer O:} $(\alpha_{2000},\delta_{\rm
  2000})=(10^h45^m,-60\degree 26.5\arcmin)$. Identified recently by
SP10, Spitzer O, located in the extreme southeast corner of the
South Pillars, is a far outpost of massive star formation in the
Carina complex (Figure~\ref{gallery}{\it d}). The cluster is associated with
the southern of two
C$^{18}$O cores in the molecular cloud known as the Giant
Pillar (Figure~\ref{image}). Although it is near the edge of the
CCCP field, where the ACIS 
sensitivity is lower, Spitzer O is nevertheless associated with a
surface density enhancement in the X-ray sources
(Figure~\ref{newimage}), which was not 
listed by F11 on account of its remote
location. Spitzer O includes a small IR dark cloud that hosts one of
the few examples of an EGO found in the Carina Nebula (SP10). This EGO
appears to be associated with 4 Stage 0/I YSOs (Table~\ref{mod}) in a
tight, linear grouping. Extending this line westward along the IR dark cloud (to
the right in Figure~\ref{gallery}{\it d}), we find a Stage II YSO and a candidate
low-luminosity ultracompact \hii region, suggesting a 
time-sequence of intermediate-mass star formation. Further to the
west, Spitzer O also
contains at least one massive star inside a compact \hii region,
identifiable as a small bubble outlined in 8~\um\ emission containing
diffuse 24~\um\ emission from dust heated by the central ionizing
source, most likely a late O-type 
star (P11). Two similar but slightly less bright \hii\ regions,
complete with their own ionizing stars, are visible to the north and
south of cluster O. All 3 candidate OB stars in Figure~\ref{gallery}{\it d} are
detected in X-rays, and none appears to have been previously cataloged
(P11).

\section{Discussion}

\subsection{A Recent Star Formation History for the Carina Nebula}

In \S\ref{YMF_SFR}, we derived a lower limit on the recent, global SFR
in the Carina Nebula of
${>}0.008$ \Msun~yr$^{-1}$, averaged over the past $\tau_0\la$~2 Myr,
traced by the IR excess sources included in 
the PCYC.
Feedback from massive stellar winds and radiation has been invoked
by many authors as the driver of ongoing star formation 
\citep[e.g.,][SP10]{NS00,SB07}. 
If triggering by massive stars is indeed the dominant mode of
continuing star formation in GMCs
that host energetic \hii regions, we can assess
its relative importance 
by comparing the present-day SFR to the historical
SFR that produced the massive star clusters that ionize the Carina
Nebula and provide the source of the feedback.

\citet{SB07} estimated a total stellar population of 5--$8\times
10^4$ stars down to the
hydrogen-burning limit by extrapolating from the
known O-type stars in Carina \citep{NS06a}.
The relative rarity of O stars requires an
enormous extrapolation over the IMF to reach the low-mass stars that
contain the bulk of the mass \citep{Kroupa}.
Measurements of the IMF among low-mass stellar populations (from
the Galactic field and open clusters) and high-mass stellar
populations (from very massive Galactic and extragalactic clusters)
have generally found slopes 
consistent with the Salpeter--Kroupa value, but
region-to-region variations 
do exist \citep{IMF10}. Because very massive
clusters with a well-populated upper IMF are located at such large
distances that individual low-mass stars generally cannot be resolved,
there has been no simultaneous,
self-consistent measurement of the IMF
from solar-mass stars to the most massive stars. Compounding the
problem, theoretical calibrations of the masses of O stars encounter a
factor of ${\sim}2$ discrepancy between ``spectroscopic'' masses derived from
surface gravity and ``evolutionary'' masses expected from stellar
structure \citep{MSH05,WV10}. Finally, in the specific case of the Carina
Nebula the completeness of the sample of known O stars \citep{NS06a}
may be as low 
as 50\% (P11).


The CCCP X-ray catalog itself offers an independent measure of the
total stellar population. The X-ray luminosity function (XLF) has
proven to be a reliable tracer of the IMF \citep{KG06} that is most
sensitive to low- and intermediate-mass PMS stars. F11 scale the XLF
from the {\it Chandra} Orion Ultradeep Project
\citep[COUP;][]{COUPglobal}, for which the underlying IMF is
well-characterized \citep{Muench}, to the XLFs from 4 distinct broad
spatial regions within the CCCP and predict a total population of
${\sim}10^5$ stars above the brown dwarf mass limit.
Because X-ray detection efficiencies are lower for YSOs versus
diskless PMS stars \citep{TP05} and for heavily obscured versus
lightly obscured stars \citep[e.g.\ Figure~3 in][]{COUPglobal}, the
simple XLF scaling analysis implicitly assumes that any differences in
the global disk fractions or the distributions of X-ray absorption
between COUP and CCCP are negligible. 

%
\begin{figure}
\vspace{1cm}
\plotone{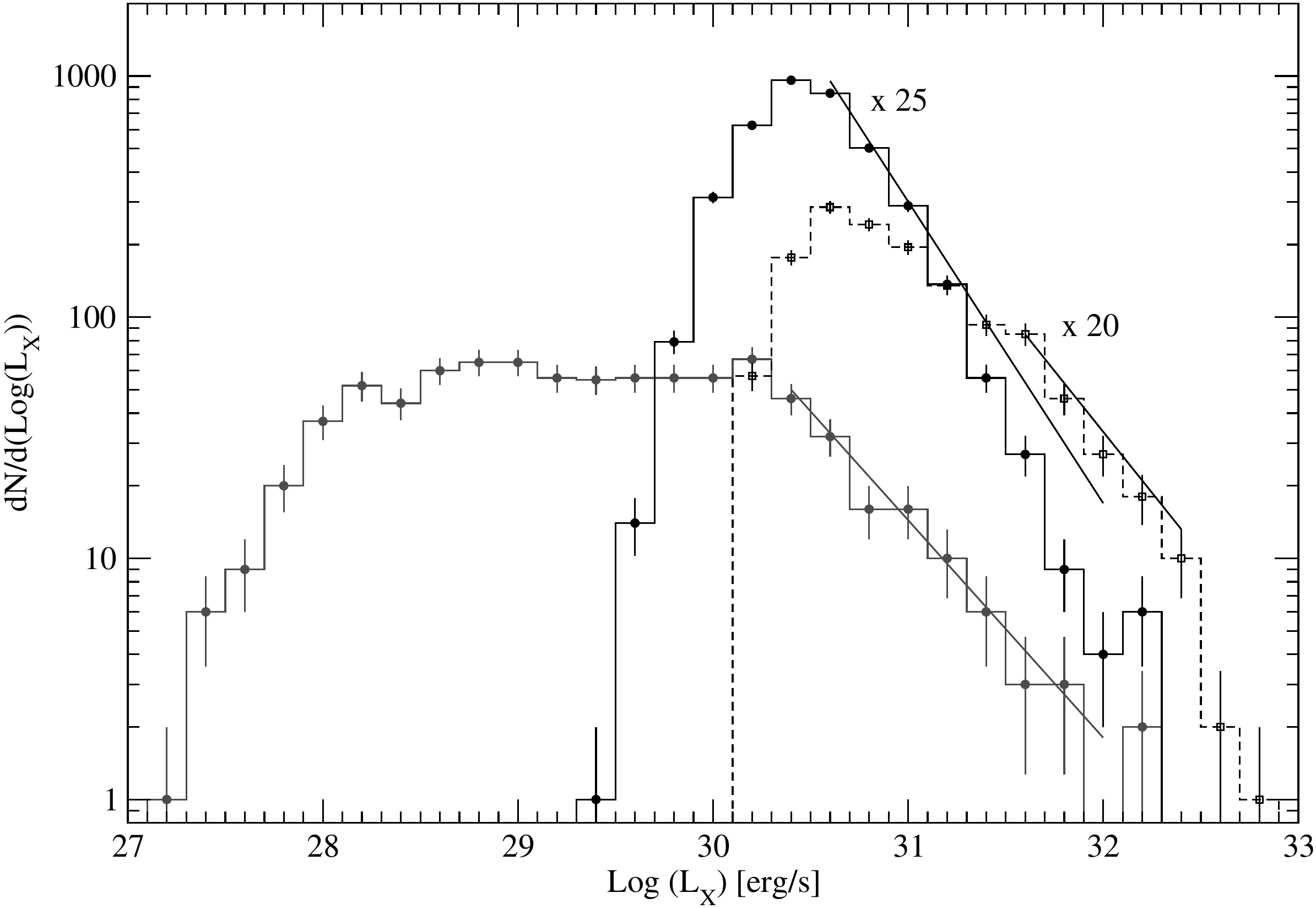}
\caption{Comparison of global total-band (0.5--8~keV),
  absorption-corrected X-ray luminosity 
  functions (XLFs) for diskless 
  candidate PMS stars in the CCCP to the XLF (gray histogram) of 839
  lightly obscured 
  cool stars from the COUP sample \citep[][]{COUPglobal}. XLFs are
  analyzed separately for the lightly-obscured ($E_{\rm median}\la
  2$~keV; solid black histogram) and heavily-obscured ($E_{\rm
    median}>2$~keV; dashed histogram) CCCP samples, and the scale
  factors used to match the COUP XLF to each CCCP XLF are labeled on
  the plot. Best-fit (negative) power-law slopes to each XLF are overplotted as
  lines; these are $\Gamma_{\rm XLF}=0.9$, $1.3$, and $1.0$ for the
  COUP, lightly-obscured CCCP, and heavily-obscured CCCP XLFs, respectively.
\label{XLF}}
\end{figure}

In Figure~\ref{XLF} we present a more detailed analysis of the CCCP
XLF. Two XLFs, separated according to median energy into
lightly-obscured ($E_{\rm median}\le 2$~keV) and heavily-obscured
($E_{\rm median}>2$~keV) samples, were constructed using all 
CCCP sources identified as 
cluster members that were sufficiently bright
to enable the derivation of absorption-corrected luminosities $L_X$
\citep{KG10,CCCPclassifier}. 
The 410 CCCP sources matched to YSOs in the PCYC were excluded, as
were ${\sim}200$ X-ray detected known or candidate OB stars (P11). The
resulting XLFs (Figure~\ref{XLF}) are clean samples of diskless PMS
stars that can be more directly compared to the 839 lightly obscured
stars in the COUP XLF, the vast majority of which lack circumstellar
disks \citep{COUPglobal}. The scale factors between the COUP and CCCP
XLFs are 25 and 20, for the lightly and heavily obscured samples,
respectively, giving a total of $45\times 839 = 3.8\times 10^4$
diskless CCCP cluster members with masses ${>}0.1$~\Msun. This result
is a lower limit due to the assumption that sources for which $L_X$
could not be obtained actually have $L_X$ values below the
XLF turnoffs at $\log{(L_X/[{\rm erg~s^{-1}}])}\sim30.6$ and 31.6 for
the lightly and heavily obscured samples, respectively (Figure~\ref{XLF}).

Combining ${\ga}3.8\times 10^4$ diskless stars traced
by the XLF (Figure~\ref{XLF}) with ${\ga}2\times 10^4$ disk-bearing
YSOs predicted by the YMF (Figure~\ref{YMF}) gives ${\ga}5.8\times 10^4$ young
stars for the global Carina Nebula population, in agreement with the
estimates of F11 and \citet{SB07}. The total stellar mass is
4.9--$8.6\times 10^4$~\Msun\ (the upper bound is computed using the XLF
scaling analysis from F11).
Because the large majority of CCCP
sources classified as likely Carina members \citep{CCCPclassifier} are
associated with the clustered stellar population (F11), we deduce that
the age spread represented by the global XLF is similar to the
${\sim}5$~Myr age
spreads measured for the major ionizing clusters \citep[][and
references therein]{CCCPTr15,CCCPTr16}. The time-averaged global SFR has been
0.010--0.017~\Msun~yr$^{-1}$ over the past ${\sim}5$~Myr. 


The combined YMF and XLF results above lead straightforwardly to a
{\it global} mid-IR excess fraction of ${\sim}30\%$
among the Carina Nebula population. Compared to the commonly-used
$K_S$-excess fraction, this mid-IR excess fraction is much closer to
the astrophysical  
fraction of stars with inner disks. YSOs dominated by circumstellar 
disks or envelopes do {\it not} separate cleanly from reddened, diskless
stars in $JHK_S$ color-color diagrams \citep[][RW06]{BW03b}.
Applying the $K_S$-excess selection criteria of \citet{CCCPhawki} to the PCYC
sources detected in 2MASS $JHK_S$ (Table~\ref{obs}), we find that only
36\% of mid-IR selected YSOs exhibit a $K_S$ excess. 
We thus predict a global $K_S$ excess fraction among the Carina Nebula
population of $30\%\times 36\%=11\%$. This result agrees very well with the
${\la}10$\% global $K_{S}$-excess fraction
reported by \citet{CCCPhawki} from deep $JHK_S$ imaging of the CCCP
stellar population, considering that the X-ray detection frequency is
expected to be lower for
YSOs than for diskless PMS stars.
We note that the stellar population in the Carina Nebula is far from
coeval, hence large variations from the ${\sim}30\%$ global average
mid-IR excess fraction are expected among different sub-populations,
for example the
relatively old Tr 15 cluster \citep{FMF80} has a very low disk
fraction \citep[see Figure~\ref{newimage} and also][]{CCCPTr15} while the very
young Treasure Chest 
cluster may have a disk 
fraction near $100\%$ \citep{TC}. The disk fraction also
decreases as a function of increasing stellar mass (PW10); the global value is
dominated by the numerous low-mass stars, while the OB stellar
population generally exhibits a low frequency of marginal mid-IR
excess emission that likely is {\it not} indicative of circumstellar disks
(P11). 


In the South Pillars region, SP10 observed a general pattern
of YSOs distributed spatially in the cavities between the ends of pillars and
the ionizing stars
and concluded that when
star formation is triggered by feedback from
massive stars, YSOs are left in the wake of retreating,
evaporating pillars and molecular cloud surfaces eroded by advancing
ionization fronts. Just over 50\% of the YSOs in the PCYC are located
in the South Pillars field 
presented by SP10 (Figures~\ref{image} and \ref{newimage}). The
wide-field coverage of the Vela--Carina survey 
allows us to generalize this picture of ongoing, potentially
feedback-driven star 
formation beyond the South Pillars to the entire Carina
Nebula. If triggering were indeed the dominant mode for ongoing star formation,
then the high present-day SFR in the Carina Nebula would demonstrate
that feedback can provide the mechanism for sustaining and propagating
star formation throughout a GMC complex.

We caution, however, that while most of the active star formation has
migrated outward \citep{SB07}, making a clear division
between a triggered YSO population and the OB stars that
provide the trigger would be an oversimplification. 
The large concentrations of YSOs associated with
the central, massive Tr 16 and Tr 14 clusters (Figures~\ref{image} and
\ref{newimage}) indicate that
these clusters are hotbeds of recent star formation; they are still
being built.
In contrast, the older Tr 15 and Bochum 11
clusters, both located in the outer regions of the Carina Nebula, have
no associated YSO sub-clusters even though 
both are obvious X-ray star clusters (Figure~\ref{newimage}).
The Carina Nebula is famous for its evolved super-massive stellar
population, $\eta$~Car being the capstone, but the numerous OB stars
apparently associated with compact clusters or groups of YSOs
(Figure~\ref{gallery}, and see also Figure~5 of P11) could be as
young as a few $10^5$~yr, especially
considering that the intense winds and radiation fields of massive stars
likely destroy their natal envelopes over very short timescales
compared to the disk lifetimes of lower-mass stars \citep{DH94}.

The triggered star formation hypothesis is very difficult to test in
a region as complicated as Carina; proof would require additional
information beyond the observed correlation between YSOs and
ionization fronts to establish a causal relationship that the ongoing
star formation {\it could not have occurred in the
  absence of a triggering mechanism provided by the massive
  stars}. High-resolution interferometric radio or sub-mm detections of dense,
massive, molecular cores
that lack active star formation and remain protected from external
radiation fields within the Carina molecular clouds could provide such
evidence. 

An alternate hypothesis of turbulence-driven, hierarchical fragmentation and
collapse of the Carina GMC complex \citep[][and references
therein]{ES07} also appears consistent with 
currently available data. 
This idea could explain the cluster-of-clusters morphology of the
Carina stellar populations and is supported by our results that a large,
approximately coeval YSO 
population is widely distributed throughout the Carina complex and
the present-day SFR is approximately the same as the time-averaged
historical SFR. In this scenario, feedback from massive stars need not
{\it trigger} star formation, but it could {\it regulate} star
formation by rapidly
sweeping the natal molecular cloud fragments away from young embedded clusters,
revealing and even unbinding these YSO clusters (as suggested by SP10)
over shorter 
timescales than expected from observations of more pristine molecular cloud
environments such as M17 SWex (PW10).


In summary, the global star formation history of the Carina GMC
complex is characterized by an approximately constant SFR of 
0.010--0.017~\Msun~yr$^{-1}$ over the past ${\ga}5$~Myr, perhaps set
in motion by the hierarchical collapse and fragmentation of the
GMC into a cluster of clusters. This average
SFR has been punctuated by
intense bursts of massive star formation. The largest burst
produced the super-massive stellar population in Tr 16, including
$\eta$~Car, about 3--4~Myr ago \citep[e.g.][]{DGE01,CCCPTr16},  
followed by a second burst that formed the very massive stellar population
in the core of the Tr 14 cluster, perhaps as recently as 0.5~Myr ago
\citep{HS10}. Subsequent to the formation of the very massive ionizing
clusters, the recent star formation history in the Carina Nebula has
been driven or at least regulated by feedback from the massive stars.

\subsection{The Production of X-rays by Intermediate-Mass PMS Stars}

Our comparative study in \S\ref{XYSOs} between the subpopulations of
the PCYC with and 
without X-ray matches in the CCCP catalog \citep{CCCPcatalog} reveals
some qualitative trends that provide clues into the origin of X-ray
emission in intermediate-mass PMS stars.
Across the ${\la}1$--10~\Msun\ mass range sampled by the PCYC
(Figure~\ref{YMF_ACIS}), X-ray emission appears to be activated during
the protostellar evolutionary phase, and this emission becomes 
more readily detectable in relatively evolved Stage 0/I objects for which the
absorption to the central star is reduced. It is well-established that
the X-ray luminosity of low-mass, T Tauri stars increases as they evolve from
disk-dominated (Stage/Class II) to diskless (Class III) PMS stars
\citep{KS04,XEST,COUPprotostars}. 
In stark contrast, we find that intermediate-mass,
disk-bearing YSOs  
continue to produce X-rays as they evolve beyond the Stage 0/I phase,
but this emission fades away as the circumstellar disks evolve 
and the central stars pass through a Herbig Ae/Be phase during their
approach to the main sequence.

X-ray detections of intermediate-mass, disk-dominated
YSOs in the mass range 1.6--4~\Msun\ are correlated with higher disk mass
{\it only} for stars with {\it cool} photospheric temperatures
(Figure~\ref{SII_comp}). 
Although these stars will arrive on the main
sequence as A and B stars with $T_{\rm eff}\ge 7500$~K and fully
radiative envelopes, currently
they have $T_{\rm eff}<5500$~K and spectral types later than
G2. During their early PMS evolution, intermediate-mass stars can thus
be regarded, in terms of their stellar structure and surface
temperatures, as more luminous 
analogs of classical T Tauri stars, complete with convective
envelopes. We therefore suggest that X-ray emission from
intermediate-mass YSOs originates in a scaled-up version of the
mechanism powering X-rays from T Tauri stars;
convection-driven magnetic reconnection activity
 \citep{KS04,TP05}. Because intermediate-mass
 stars reach the main sequence within a few Myr \citep{SDF00},
 this emission mechanism is available for only a short time compared to
 low-mass PMS stars. The rapid fading of X-ray emission coincides with the
rapid dissipation of the circumstellar disk as the temperature of the
central star rises.


This rapidly fading dynamo hypothesis can also explain a puzzling
feature of the CCCP XLF. The slope of the lightly-obscured CCCP XLF,
fitted with a power-law function using the method
of \citet{powerlaw09}, is $\Gamma_{\rm XLF}=1.3$, significantly steeper
than the slopes of both the the comparison lightly-obscured COUP XLF
($\Gamma_{\rm XLF}=0.9$) and the heavily-obscured CCCP XLF
($\Gamma_{\rm XLF}=1.0$; Figure~\ref{XLF}). If the lightly-obscured
CCCP XLF traces an older stellar population compared to either the
COUP or the heavily-obscured 
CCCP XLFs, then a relatively rapid decay in $L_X$ among the
more luminous, intermediate-mass stars would steepen the observed XLF
slope, just as mass-dependent circumstellar disk evolution steepens the
observed YMF slope (Figure~\ref{YMF}). A similar deficit of stars with
$\log{L_X}\ga 31~{\rm erg~s^{-1}}$ steepens the observed XLF slope of the
Tr 15 cluster \citep{CCCPTr15}. With an age ${\ga}5$~Myr, Tr 15
represents a significantly older stellar population compared to either
the ONC or the other massive clusters in the Carina Nebula.



\section{Conclusions}

We have used the \spitzer\ Vela--Carina point-source lists, in
combination with complementary IR datasets, to carry out an unbiased
search for YSOs with IR excess emission throughout the Carina
Nebula. We produce a Pan-Carina YSO Catalog (PCYC) of 1439
highly-reliable candidate YSOs, 410 of which have X-ray counterparts from the
CCCP catalog. The PCYC comprises one of the largest samples of
intermediate-mass YSOs in a single massive star-forming complex
compiled to date, and it is the largest such collection with a
complete set of complementary X-ray observations. 
From our initial analysis of this rich dataset, we can draw the
following conclusions:
\begin{itemize}
  \item The mass function (YMF) of the Carina YSOs exhibits a
    power-law slope $\Gamma_{\rm YMF}=3.3$, significantly
    steeper than the Salpeter--Kroupa IMF. 
    Such a steep slope, first observed
    for the YSO population of the M17 SWex IR dark cloud,
    is at least partially the result of an observational selection effect
    caused by rapid disk  
    evolution among intermediate-mass YSOs 
    (PW10). 
    M17 SWex and Carina are very different environments; the
    former is a cold, IR dark cloud apparently experiencing its first wave of
    star formation while the latter is an evolved \hii region powered
    by some of the 
    most massive stars in the Galaxy.
    While the YMF slopes are similar, the lower critical mass 
    of the PCYC sample, below which the YMF may agree with a normal
    Salpeter--Kroupa IMF, likely reflects a larger age spread among the
    YSOs in the Carina Nebula.
  \item The time-averaged global SFR in the Carina complex has been
    0.010--0.017~Msun~yr$^{-1}$ over the past ${<}5$~Myr,
    punctuated by intense bursts of activity that produced the very
    massive stellar populations in the Tr 16 and Tr 14
    clusters. If the present-day Galactic SFR is a few
    \Msun~yr$^{-1}$, then the star formation activity of the
    entire Milky Way is equivalent to a few hundred Carina complexes.
    The cluster-of-clusters pattern of star formation in the Carina Nebula is
    consistent with
    hierarchical fragmentation and collapse of a GMC complex \citep{ES07}.
  \item The present-day global star formation rate in the Carina Complex is
    ${>}0.008$~\Msun~yr$^{-1}$. The 
    massive central ionizing clusters Tr 16 and Tr 14 still host
    intense concentrations of ongoing star formation. 
    Ongoing star
    formation activity is distributed widely throughout Carina,
    driven or regulated by feedback from the massive stellar
    population, which acts to quickly reveal and even unbind young,
    embedded clusters (SP10). 
  \item The PCYC includes 62 candidate X-ray-emitting Stage 0/I YSOs, a
    sample large enough to provide statistical evidence that X-ray
    detections correlate with more evolved protostars. Our results support
    previous findings that the physical mechanism producing X-rays in
    PMS stars is activated during the protostellar
    evolutionary phase, and circumstellar material contributes
    significantly to the absorption of X-rays from the youngest YSOs
    \citep{KG07,COUPprotostars}. 
  \item The apparent X-ray emission from intermediate-mass
    YSOs is usually consistent with the presence of unresolved,
    low-mass companions. But we also isolate a class of intermediate-mass
    Stage II YSOs with cooler photospheres and relatively massive
    circumstellar disks for which the X-ray emission appears to be
    intrinsic. We suggest that the progenitors of Herbig Ae/Be stars
    produce X-rays during their early PMS evolution, perhaps powered
    by magnetic reconnection activity
    during a convective atmosphere phase, but this mechanism 
    decays as the stars approach the main sequence. A rapid fading of
    the stellar dynamo could thus act to steepen observed
    XLF slopes as a function of increasing cluster age \citep{CCCPTr15}.
  \item We identify 3 new very young, tight YSO sub-clusters in the
    northwestern region of the Carina Nebula, 2 of which are
    associated with significant concentrations of X-ray sources (F11).
\end{itemize}

The full PCYC dataset is available in the electronic versions
of Tables~\ref{obs} and \ref{mod}. We hope that this catalog will serve
as a valuable database for 
future multiwavelength observations and modeling of star formation in
the Carina Complex. 

\acknowledgments
We thank E. D. Feigelson, T. Preibisch, and
T. Montmerle for insightful conversations that helped improve this work.
We thank the anonymous referee for a timely and valuable review.
M.S.P. is supported by an NSF Astronomy and Astrophysics Postdoctoral
Fellowship under award AST-0901646. This work is based on
observations from the {\it Spitzer Space Telescope} GO programs 30848
(MIPSCAR; PI: N.S.) and 40791 (Vela--Carina; PI: S.R.M.). 
This publication makes use of data products from the Two Micron
All-Sky Survey, which is a joint project of the University of
Massachusetts and the Infrared Processing and Analysis
Center/California Institute of Technology, funded by NASA and the NSF.
Support for this work was provided by NASA through awards issued by
the Jet Propulsion Laboratory, California 
Institute of Technology.
This work was also supported by
{\it Chandra X-ray Observatory} grant GO-9131X (PI: L.K.T.) and by the ACIS Instrument
Team contract SV4-74018 (PI: G.\ Garmire), issued by the {\it Chandra}
X-ray Center, which is operated by the Smithsonian Astrophysical
Observatory for and on behalf of NASA under contract NAS8-03060.
K.G.S. acknowledges support from
NSF grant AST-0849736. 

Facilities: \facility{Spitzer(IRAC)}, \facility{Spitzer(MIPS)}, \facility{CTIO:2MASS}, \facility{MSX}, \facility{CXO (ACIS)}

\clearpage

\LongTables
\begin{deluxetable}{@{\hspace{-10mm}}r@{~~}c@{~~}c@{~~}c@{\hspace{-3mm}}c@{\hspace{-3mm}}c@{\hspace{-3mm}}c@{\hspace{-3mm}}c@{\hspace{-3mm}}c@{\hspace{-3mm}}c@{\hspace{-3mm}}c@{\hspace{-3mm}}c@{}c@{}c@{}c@{}c@{}c@{}c@{}c@{\hspace{-1.5mm}}cc}
\tablecaption{Pan-Carina YSO Catalog: Basic IR Observational Data\tablenotemark{a}\label{obs}}
\tablewidth{0pt}
\tablehead{
\colhead{(1)} & \colhead{(2)} & \colhead{(3)} & \colhead{(4)} & \colhead{(5)} & \colhead{(6)} & \colhead{(7)} & \colhead{(8)} & \colhead{(9)} & \colhead{(10)} & \colhead{(11)} & \colhead{(12)} & \multicolumn{8}{c}{(13)--(20)} & \colhead{(21)} \\
\colhead{PCYC} & \colhead{RA} & \colhead{Dec} & \colhead{} & \colhead{$J$} & \colhead{$H$} & \colhead{$K_S$} & \colhead{[3.6]} & \colhead{[4.5]} & \colhead{[5.8]} & \colhead{[8.0]} & \colhead{[24]} & \multicolumn{8}{c}{Quality} & \colhead{} \\
\colhead{No.} & \colhead{(J2000)} & \colhead{(J2000)} & \colhead{SSTGLM\tablenotemark{b}} & \colhead{(mag)} & \colhead{(mag)} & \colhead{(mag)} & \colhead{(mag)} & \colhead{(mag)} & \colhead{(mag)} & \colhead{(mag)} & \colhead{(mag)} & \multicolumn{8}{c}{Flags\tablenotemark{c}} & \colhead{\it MSX?}
}
\startdata
 117 &   10 42 41.56  &    -59 24 43.3 &       C G287.2040-00.5300 &  \nodata &   14.5 &   13.2 &   10.7 &   10.0 &    9.3 &    8.8 &    4.3 &  0 &  1 &  1 &  1 &  1 &  1 &  1 &  3 &            \\
 118 &   10 42 41.91  &    -59 24 52.7 &       C G287.2059-00.5320 &   14.2 &   13.8 &   13.6 &   13.2 &   12.6 &   11.5 &  \nodata &    5.3 &  1 &  1 &  1 &  1 &  1 &  1 &  0 &  1 &            \\
 119 &   10 42 43.34  &    -59 25 06.5 &       C G287.2104-00.5339 &  \nodata &   14.6 &   13.7 &   12.4 &   11.9 &   11.4 &   10.2 &    6.0 &  0 &  1 &  1 &  1 &  1 &  1 &  1 &  3 &            \\
 122 &   10 42 44.22  &    -59 25 20.8 &       C G287.2139-00.5365 &   15.9 &   14.5 &   13.8 &   12.8 &   12.4 &   11.6 &   10.5 &    4.6 &  1 &  1 &  1 &  1 &  1 &  1 &  1 &  3 &            \\
 124 &   10 42 45.36  &    -59 25 25.9 &       C G287.2167-00.5366 &  \nodata &  \nodata &   14.3 &   12.7 &   12.2 &   12.0 &   10.6 &    4.6 &  0 &  0 &  1 &  1 &  1 &  1 &  1 &  3 &            \\
 125 &   10 42 45.38  &    -59 25 13.7 &       C G287.2151-00.5336 &   15.4 &   14.0 &   13.2 &   12.1 &   11.6 &   11.4 &   10.2 &    4.4 &  1 &  1 &  1 &  1 &  1 &  1 &  1 &  3 &            \\
 126 &   10 42 45.40  &    -59 25 39.2 &       C G287.2185-00.5399 &  \nodata &  \nodata &  \nodata &   13.2 &   12.5 &   11.8 &   10.8 &    4.8 &  0 &  0 &  0 &  1 &  1 &  1 &  1 &  3 &            \\
 129 &   10 42 46.29  &    -59 25 31.4 &       C G287.2192-00.5370 &  \nodata &  \nodata &  \nodata &   13.7 &   11.7 &   10.8 &   10.3 &    3.0 &  0 &  0 &  0 &  1 &  1 &  1 &  1 &  1 &            \\
 133 &   10 42 46.95  &    -59 25 20.9 &       C G287.2190-00.5338 &  \nodata &  \nodata &   15.0 &   13.8 &   12.2 &   11.3 &  \nodata &    2.9 &  0 &  0 &  1 &  1 &  1 &  1 &  0 &  1 &            \\
 138 &   10 42 48.09  &    -59 25 39.4 &       C G287.2236-00.5372 &  \nodata &   14.1 &   13.2 &   12.2 &   11.7 &   11.3 &   10.5 &    2.8 &  0 &  1 &  1 &  1 &  1 &  1 &  1 &  1 &            \\
 139 &   10 42 48.14  &    -59 25 29.0 &       C G287.2223-00.5346 &   12.6 &   11.6 &   10.1 &    7.6 &    6.9 &    6.0 &    5.1 &  \nodata &  1 &  1 &  1 &  1 &  1 &  1 &  1 &  0 &            \\
 141 &   10 42 48.58  &    -59 25 38.3 &       C G287.2244-00.5364 &  \nodata &  \nodata &   14.4 &   12.1 &   11.3 &   10.6 &   10.0 &    3.0 &  0 &  0 &  1 &  1 &  1 &  1 &  1 &  1 &            \\
 144 &   10 42 49.74  &    -59 25 36.5 &       C G287.2263-00.5348 &  \nodata &   14.4 &   13.6 &   11.7 &   11.1 &   10.7 &    9.9 &    4.1 &  0 &  1 &  1 &  1 &  1 &  1 &  1 &  3 &            \\
 145 &   10 42 50.47  &    -59 25 43.8 &       C G287.2286-00.5359 &   15.6 &   13.5 &   11.8 &    9.8 &    9.1 &    8.5 &    7.7 &    4.1 &  1 &  1 &  1 &  1 &  1 &  1 &  1 &  3 &            \\
 147 &   10 42 50.62  &    -59 25 25.4 &       C G287.2265-00.5312 &   14.4 &   11.6 &    9.5 &    7.4 &    6.7 &    6.1 &    5.1 &    2.0 &  1 &  1 &  1 &  1 &  1 &  1 &  1 &  1 &            \\
 148 &   10 42 51.30  &    -59 25 36.0 &       C G287.2291-00.5331 &   15.7 &   14.7 &   14.0 &   13.3 &   12.6 &   11.6 &   10.3 &    4.5 &  1 &  1 &  1 &  1 &  1 &  1 &  1 &  3 &            \\
 155 &   10 42 54.93  &    -59 25 46.2 &       C G287.2372-00.5320 &   15.6 &   14.2 &   13.3 &   12.4 &   12.0 &   11.7 &   10.6 &    5.1 &  1 &  1 &  1 &  1 &  1 &  1 &  1 &  3 &            \\
 156 &   10 42 55.41  &    -59 24 56.5 &       C G287.2316-00.5194 &   16.6 &   15.4 &   14.4 &   12.8 &   12.3 &   11.6 &   10.9 &    4.5 &  1 &  1 &  1 &  1 &  1 &  1 &  1 &  3 &            \\
 159 &   10 42 58.43  &    -59 24 52.6 &       C G287.2367-00.5154 &   14.8 &   13.3 &   12.7 &   12.1 &   12.2 &   11.7 &   11.2 &    4.6 &  1 &  1 &  1 &  1 &  1 &  1 &  1 &  3 &            \\
 161 &   10 42 58.67  &    -59 25 34.9 &       C G287.2427-00.5255 &  \nodata &  \nodata &  \nodata &   13.2 &   12.3 &   11.5 &   10.8 &    4.5 &  0 &  0 &  0 &  1 &  1 &  1 &  1 &  3 &            \\
 162 &   10 42 58.78  &    -59 25 04.1 &       C G287.2389-00.5178 &  \nodata &  \nodata &  \nodata &   12.6 &   12.0 &   11.3 &   10.0 &    4.7 &  0 &  0 &  0 &  1 &  1 &  1 &  1 &  3 &            \\
 164 &   10 43 00.32  &    -59 25 27.2 &       C G287.2448-00.5220 &   16.0 &   14.7 &   13.7 &   12.2 &   11.8 &   11.5 &   10.7 &    4.4 &  1 &  1 &  1 &  1 &  1 &  1 &  1 &  3 &            \\
\hline
 177 &   10 43 07.36  &    -59 24 33.4 &       C G287.2509-00.5017 &   15.8 &   13.9 &   12.5 &   10.8 &   10.3 &    9.9 &    9.4 &    4.8 &  1 &  1 &  1 &  1 &  1 &  1 &  1 &  3 &            \\
 179 &   10 43 09.72  &    -59 24 55.3 &       C G287.2582-00.5047 &  \nodata &  \nodata &  \nodata &   11.3 &    9.0 &    7.0 &    5.9 &    2.2 &  0 &  0 &  0 &  1 &  1 &  1 &  1 &  1 & \checkmark \\
 183 &   10 43 11.37  &    -59 25 06.1 &       C G287.2627-00.5057 &   15.7 &   14.3 &   13.8 &   12.7 &   12.4 &   11.8 &   11.2 &    4.9 &  1 &  1 &  1 &  1 &  1 &  1 &  1 &  3 &            \\
 184 &   10 43 11.72  &    -59 24 39.3 &       C G287.2598-00.4988 &  \nodata &  \nodata &  \nodata &   14.8 &   12.9 &   11.8 &   10.9 &    5.5 &  0 &  0 &  0 &  1 &  1 &  1 &  1 &  3 &            \\
 185 &   10 43 11.91  &    -59 24 52.7 &       C G287.2619-00.5019 &  \nodata &  \nodata &  \nodata &   13.7 &   12.4 &   11.6 &   11.0 &    4.9 &  0 &  0 &  0 &  1 &  1 &  1 &  1 &  3 &            \\
 186 &   10 43 12.06  &    -59 24 58.8 &       C G287.2630-00.5032 &   15.3 &   13.9 &   13.2 &   12.0 &   11.6 &   10.9 &   10.2 &    4.8 &  1 &  1 &  1 &  1 &  1 &  1 &  1 &  3 &            \\
 187 &   10 43 12.22  &    -59 24 18.1 &       C G287.2580-00.4931 &  \nodata &  \nodata &  \nodata &   13.6 &   12.6 &   11.6 &   10.4 &    3.7 &  0 &  0 &  0 &  1 &  1 &  1 &  1 &  3 &            \\
 189 &   10 43 12.62  &    -59 25 15.1 &       C G287.2662-00.5067 &   16.1 &   14.7 &  \nodata &   12.6 &   12.1 &   11.8 &   11.3 &    3.6 &  1 &  1 &  0 &  1 &  1 &  1 &  1 &  3 &            \\
 192 &   10 43 14.00  &    -59 23 53.8 &       C G287.2581-00.4854 &   14.3 &   13.5 &   13.0 &   12.0 &   11.6 &   11.1 &   10.4 &    4.4 &  1 &  1 &  1 &  1 &  1 &  1 &  1 &  3 &            \\
 194 &   10 43 14.24  &    -59 25 09.0 &       C G287.2684-00.5035 &   16.5 &   15.2 &   14.3 &   13.2 &   12.6 &   12.1 &   11.2 &    3.7 &  1 &  1 &  1 &  1 &  1 &  1 &  1 &  3 &            \\
 197 &   10 43 14.66  &    -59 24 23.4 &       C G287.2632-00.4920 &   16.4 &   15.2 &   14.6 &   13.6 &   13.0 &   12.1 &  \nodata &    4.7 &  1 &  1 &  1 &  1 &  1 &  1 &  0 &  3 &            \\
\hline
 372 &   10 43 55.75  &    -59 08 48.3 &       C G287.2180-00.2216 &  \nodata &  \nodata &  \nodata &   13.0 &   12.1 &   11.4 &   10.6 &    6.5 &  0 &  0 &  0 &  1 &  1 &  1 &  1 &  3 &            \\
 391 &   10 43 58.38  &    -59 08 43.3 &       C G287.2223-00.2178 &  \nodata &  \nodata &   14.6 &   13.5 &   12.1 &   11.2 &   10.4 &    5.6 &  0 &  0 &  1 &  1 &  1 &  1 &  1 &  3 &            \\
 400 &   10 44 00.23  &    -59 08 54.3 &       C G287.2272-00.2186 &  \nodata &  \nodata &  \nodata &   12.6 &   10.7 &    9.3 &    8.0 &    3.8 &  0 &  0 &  0 &  1 &  1 &  1 &  1 &  1 &            \\
 407 &   10 44 00.85  &    -59 08 56.3 &       C G287.2287-00.2185 &  \nodata &  \nodata &  \nodata &   13.2 &   12.2 &   10.9 &    9.9 &    5.1 &  0 &  0 &  0 &  1 &  1 &  1 &  1 &  1 &            \\
 416 &   10 44 02.55  &    -59 08 24.5 &       C G287.2277-00.2090 &  \nodata &  \nodata &  \nodata &   12.8 &   12.3 &   11.3 &   10.5 &    6.8 &  0 &  0 &  0 &  1 &  1 &  1 &  1 &  3 &            \\
 417 &   10 44 02.77  &    -59 08 49.4 &       C G287.2314-00.2148 &  \nodata &  \nodata &  \nodata &   13.1 &   11.5 &   10.5 &    9.8 &    5.4 &  0 &  0 &  0 &  1 &  1 &  1 &  1 &  3 &            \\
 426 &   10 44 04.89  &    -59 07 56.5 &       C G287.2285-00.1998 &   13.6 &   13.0 &   12.8 &   12.7 &   12.6 &   12.0 &   10.9 &    6.2 &  1 &  1 &  1 &  1 &  1 &  1 &  1 &  3 &            \\
 428 &   10 44 04.96  &    -59 08 33.3 &       C G287.2334-00.2087 &  \nodata &   15.4 &   14.3 &   13.0 &   12.6 &   12.0 &   11.4 &    7.5 &  0 &  1 &  1 &  1 &  1 &  1 &  1 &  3 &            \\
 438 &   10 44 07.04  &    -59 08 39.2 &       C G287.2381-00.2081 &  \nodata &   14.1 &   12.6 &   10.6 &    9.4 &    7.9 &    6.6 &    2.5 &  0 &  1 &  1 &  1 &  1 &  1 &  1 &  1 &            \\
 442 &   10 44 08.12  &    -59 08 11.2 &       C G287.2365-00.2001 &  \nodata &   15.4 &   14.4 &   13.3 &   12.8 &   11.8 &  \nodata &    5.8 &  0 &  1 &  1 &  1 &  1 &  1 &  0 &  3 &            \\
\hline
 528 &   10 44 25.99  &    -59 33 38.8 &       C G287.4690-00.5571 &   13.6 &   13.0 &   12.6 &   11.9 &   11.6 &   11.4 &   10.8 &  \nodata &  1 &  1 &  1 &  1 &  1 &  1 &  1 &  0 &            \\
 556 &   10 44 31.11  &    -59 33 10.1 &       C G287.4748-00.5450 &  \nodata &  \nodata &  \nodata &    8.8 &    7.4 &    6.0 &    5.0 &  \nodata &  0 &  0 &  0 &  1 &  1 &  1 &  1 &  0 &            \\
 563 &   10 44 32.21  &    -59 33 59.6 &       C G287.4833-00.5561 &   11.0 &   10.8 &   10.8 &   10.7 &   10.7 &   10.4 &    9.6 &  \nodata &  1 &  1 &  1 &  1 &  1 &  1 &  1 &  0 &            \\
 574 &   10 44 33.07  &    -59 34 33.2 &       C G287.4893-00.5635 &  \nodata &  \nodata &  \nodata &   13.6 &   13.6 &   11.4 &   10.0 &  \nodata &  0 &  0 &  0 &  1 &  1 &  1 &  1 &  0 &            \\
 575 &   10 44 33.18  &    -59 33 49.1 &       C G287.4837-00.5526 &   13.4 &   12.1 &   11.3 &   10.1 &    9.7 &    9.3 &    8.7 &  \nodata &  1 &  1 &  1 &  1 &  1 &  1 &  1 &  0 &            \\
 600 &   10 44 37.01  &    -59 34 32.5 &       C G287.4965-00.5594 &   12.5 &   12.1 &   11.6 &   10.7 &   10.1 &    9.7 &    8.9 &  \nodata &  1 &  1 &  1 &  1 &  1 &  1 &  1 &  0 &            \\
 612 &   10 44 38.82  &    -59 33 47.2 &       C G287.4940-00.5465 &   12.0 &   10.9 &  \nodata &    8.1 &    7.6 &    7.2 &    6.5 &  \nodata &  1 &  1 &  0 &  1 &  1 &  1 &  1 &  0 &            \\
 616 &   10 44 39.01  &    -59 34 30.6 &       C G287.5000-00.5570 &   15.9 &   14.8 &   14.2 &   13.3 &   12.9 &   12.1 &   11.3 &  \nodata &  1 &  1 &  1 &  1 &  1 &  1 &  1 &  0 &            \\
 619 &   10 44 39.79  &    -59 32 40.5 &       C G287.4871-00.5292 &   14.7 &   13.9 &   13.6 &   12.9 &   12.4 &   11.6 &   10.3 &  \nodata &  1 &  1 &  1 &  1 &  1 &  1 &  1 &  0 &            \\
 621 &   10 44 40.03  &    -59 33 46.5 &       C G287.4962-00.5451 &  \nodata &  \nodata &   13.9 &   12.8 &   12.3 &   11.6 &  \nodata &    3.2 &  0 &  0 &  1 &  1 &  1 &  1 &  0 &  3 &            \\
 625 &   10 44 40.56  &    -59 33 54.3 &       C G287.4982-00.5465 &  \nodata &  \nodata &  \nodata &   12.8 &   12.5 &   12.0 &   11.0 &    2.8 &  0 &  0 &  0 &  1 &  1 &  1 &  1 &  3 &            \\
 634 &   10 44 41.46  &    -59 33 56.4 &       C G287.5001-00.5462 &  \nodata &  \nodata &  \nodata &   12.0 &   11.3 &   11.0 &   10.2 &    2.8 &  0 &  0 &  0 &  1 &  1 &  1 &  1 &  3 &            \\
 642 &   10 44 42.23  &    -59 33 35.6 &       C G287.4989-00.5403 &   15.2 &   14.2 &   13.3 &   12.0 &   11.4 &   11.0 &  \nodata &    4.0 &  1 &  1 &  1 &  1 &  1 &  1 &  0 &  3 &            \\
 654 &   10 44 44.05  &    -59 34 10.4 &       C G287.5068-00.5470 &  \nodata &  \nodata &  \nodata &   13.7 &   12.8 &   11.7 &   10.4 &    2.3 &  0 &  0 &  0 &  1 &  1 &  1 &  1 &  3 &            \\
 655 &   10 44 44.06  &    -59 34 26.2 &       C G287.5088-00.5509 &   15.1 &   13.1 &   11.7 &   10.2 &    9.3 &    8.4 &    7.4 &  \nodata &  1 &  1 &  1 &  1 &  1 &  1 &  1 &  0 &            \\
 656 &   10 44 44.09  &    -59 34 07.2 &       C G287.5064-00.5462 &  \nodata &  \nodata &  \nodata &   13.4 &   12.1 &   11.1 &    9.7 &    2.5 &  0 &  0 &  0 &  1 &  1 &  1 &  1 &  3 &            \\
 657 &   10 44 44.23  &    -59 34 05.1 &       C G287.5064-00.5456 &  \nodata &  \nodata &  \nodata &   12.6 &   11.5 &   10.5 &    9.6 &    2.6 &  0 &  0 &  0 &  1 &  1 &  1 &  1 &  3 &            \\
 660 &   10 44 44.41  &    -59 32 35.3 &       C G287.4951-00.5234 &   15.5 &   14.3 &   13.4 &   12.5 &   12.1 &   11.6 &  \nodata &    2.5 &  1 &  1 &  1 &  1 &  1 &  1 &  0 &  3 &            \\
\hline
 699 &   10 44 51.89  &    -60 25 11.7 &       C G287.9186-01.2913 &  \nodata &   15.0 &   14.0 &   11.9 &   10.9 &   10.1 &    9.1 &    5.3 &  0 &  1 &  1 &  1 &  1 &  1 &  1 &  1 &            \\
\hline
1295 &   10 47 33.40  &    -60 26 56.7 &       C G288.2265-01.1635 &   11.5 &   11.2 &   11.0 &   10.8 &   10.8 &   10.5 &   10.1 &    4.5 &  1 &  1 &  1 &  1 &  1 &  1 &  1 &  3 &            \\
1297 &   10 47 35.69  &    -60 26 28.5 &       C G288.2271-01.1544 &   14.5 &   13.4 &   12.6 &   11.3 &   10.7 &    9.9 &    8.8 &    4.1 &  1 &  1 &  1 &  1 &  1 &  1 &  1 &  1 &            \\
1302 &   10 47 42.09  &    -60 25 23.1 &       C G288.2305-01.1322 &   16.3 &   14.8 &   13.7 &   11.3 &   11.0 &   10.3 &    9.5 &    4.9 &  1 &  1 &  1 &  1 &  1 &  1 &  1 &  3 &            \\
1304 &   10 47 44.89  &    -60 27 26.9 &       C G288.2514-01.1602 &  \nodata &  \nodata &  \nodata &   13.6 &   12.4 &   11.5 &    9.9 &    6.1 &  0 &  0 &  0 &  1 &  1 &  1 &  1 &  3 &            \\
1308 &   10 47 47.19  &    -60 26 33.5 &       C G288.2488-01.1448 &  \nodata &  \nodata &   14.5 &   12.2 &   11.5 &   10.9 &    9.9 &    5.3 &  0 &  0 &  1 &  1 &  1 &  1 &  1 &  3 &            \\
1312 &   10 47 48.52  &    -60 26 09.5 &       C G288.2482-01.1376 &  \nodata &   14.4 &   12.7 &   10.4 &    9.7 &    9.1 &    8.3 &    5.1 &  0 &  1 &  1 &  1 &  1 &  1 &  1 &  1 &            \\
1313 &   10 47 48.97  &    -60 25 44.8 &       C G288.2458-01.1311 &  \nodata &  \nodata &   14.2 &   12.7 &   12.2 &   11.4 &   10.3 &    7.4 &  0 &  0 &  1 &  1 &  1 &  1 &  1 &  3 &            \\
1315 &   10 47 50.24  &    -60 26 18.6 &       C G288.2525-01.1383 &  \nodata &  \nodata &   13.6 &   12.3 &   11.0 &    9.9 &    9.2 &    4.3 &  0 &  0 &  1 &  1 &  1 &  1 &  1 &  1 &            \\
1316 &   10 47 50.59  &    -60 27 04.7 &       C G288.2589-01.1494 &   16.1 &   14.0 &  \nodata &   12.2 &   12.2 &   11.8 &   10.8 &    6.0 &  1 &  1 &  0 &  1 &  1 &  1 &  1 &  3 &            \\
1317 &   10 47 50.70  &    -60 26 18.1 &       C G288.2532-01.1377 &  \nodata &  \nodata &  \nodata &   12.8 &   10.8 &    9.5 &    8.6 &    4.5 &  0 &  0 &  0 &  1 &  1 &  1 &  1 &  1 &            \\
1319 &   10 47 51.67  &    -60 26 23.9 &       C G288.2557-01.1383 &  \nodata &  \nodata &   14.0 &   12.7 &   11.8 &   11.3 &   10.6 &    4.7 &  0 &  0 &  1 &  1 &  1 &  1 &  1 &  1 &            \\
1320 &   10 47 52.07  &    -60 26 27.0 &       C G288.2569-01.1386 &  \nodata &  \nodata &  \nodata &   12.2 &   10.2 &    9.0 &    8.2 &    4.8 &  0 &  0 &  0 &  1 &  1 &  1 &  1 &  1 &            
\enddata
\tablecomments{This Table is available in its entirety in 
  a machine-readable form in the online journal. A portion is
  reproduced here for guidance regarding its form and content and to
  highlight sources of interest. Sources are grouped according to
  their apparent association with the YSO sub-clusters discussed in
  \S\ref{butterflies}.}  
\tablenotetext{a}{The typical photometric uncertainty is 0.1 mag.}
\tablenotetext{b}{Source names are preceded by either a ``C'' for the highly-reliable
  Vela--Carina Catalog or an ``A'' for the more-complete Archive.}
\tablenotetext{c}{Quality flags are defined as follows: 0 =
  non-detection, 1 = detection, 2 = flux density lower limit
  (magnitude upper limit), and 3 = flux density upper limit (magnitude
lower limit).}

\end{deluxetable}

\begin{deluxetable}{rccccccccccc}
\tabletypesize{\scriptsize}
\tablecaption{Pan-Carina YSO Catalog: Modeled Properties and CCCP Matches\label{mod}}
\tablewidth{0pt}
\tablehead{
\colhead{(1)} & \colhead{(2)} & \colhead{(3)} & \colhead{(4)} & \colhead{(5)} & \colhead{(6)} & \colhead{(7)} & \colhead{(8)} & \colhead{(9)} & \colhead{(10)} & \colhead{(11)} &\colhead{(12)} \\
\colhead{PCYC} & \colhead{$\langle\log{L_{\rm bol}}\rangle$} & \colhead{$\sigma(\log{L_{\rm bol}})$} & \colhead{$\langle M_{\star}\rangle$} & \colhead{$\sigma(M_{\star})$} & \colhead{$\langle\log{\dot{M}_{\rm env}}\rangle$} & \colhead{$\sigma(\log{\dot{M}_{\rm env}})$} & \colhead{} & \colhead{CCCP Match} & \multicolumn{2}{c}{Membership\tablenotemark{a}} & \colhead{} \\
\colhead{No.} & \colhead{(\Lsun)} & \colhead{(\Lsun)} & \colhead{(\Msun)} & \colhead{(\Msun)} & \colhead{(\Msun~yr$^{-1}$)} & \colhead{(\Msun~yr$^{-1}$)} & \colhead{Stage} & \colhead{CXOGNC J} & \colhead{Cl/Gr} & \colhead{Reg} & \colhead{Notes\tablenotemark{b}}
}
\startdata
 117 &   1.9 &   0.2 &   3.4 &   1.1 &  -4.8 &   1.9 &    0/I &                           &                    &                    D &                    P \\
 118 &   1.6 &   0.1 &   2.5 &   0.2 & \nodata & \nodata &     II &                           &                    &                    D &                    P \\
 119 &   1.3 &   0.9 &   1.9 &   0.9 &  -7.1 &   2.5 &     II &                           &                    &                    D &                    P \\
 122 &   1.2 &   0.9 &   1.3 &   1.1 &  -7.3 &   2.2 &      A &                           &                    &                    D &                    P \\
 124 &   1.8 &   0.7 &   2.7 &   1.0 &  -6.3 &   2.3 &      A &                           &                   G4 &                    D &                    P \\
 125 &   1.9 &   0.5 &   2.9 &   0.8 & \nodata & \nodata &    III &        104245.34-592513.3 &                    &                    D &                 P, 3X \\
 126 &   1.7 &   1.1 &   2.4 &   1.3 &  -6.0 &   2.2 &      A &                           &                   G4 &                    D &                    P \\
 129 &   2.6 &   0.2 &   6.4 &   1.0 &  -4.0 &   0.8 &    0/I &                           &                   G4 &                    D &                    P \\
 133 &   2.3 &   0.4 &   4.0 &   1.3 &  -4.2 &   1.8 &    0/I &                           &                   G4 &                    D &                    P \\
 138 &   2.0 &   0.1 &   3.1 &   0.7 &  -4.5 &   0.3 &    0/I &        104248.10-592539.3 &                   G4 &                    D &                    P \\
 139 &   2.6 &   0.3 &   4.9 &   1.2 &  -8.3 &   1.8 &     II &        104248.17-592528.7 &                   G4 &                    D &                    P \\
 141 &   2.0 &   0.1 &   4.5 &   0.6 &  -4.4 &   2.0 &    0/I &                           &                   G4 &                    D &                    P \\
 144 &   1.3 &   0.3 &   1.6 &   0.7 &  -5.4 &   2.3 &    0/I &                           &                    &                    D &                    P \\
 145 &   2.9 &   1.7 &   4.4 &   1.9 & \nodata & \nodata &     II &        104250.50-592544.2 &                    &                    D &                    P \\
 147 &   3.1 &   0.1 &   6.5 &   0.4 & \nodata & \nodata &     II &                           &                    &                    D &                    P \\
 148 &   0.7 &   0.5 &   0.6 &   0.5 &  -4.8 &   2.2 &    0/I &        104251.41-592535.6 &                    &                    D &                    P \\
 155 &   1.7 &   1.0 &   2.3 &   1.0 &  -7.2 &   2.6 &      A &        104254.93-592545.4 &                    &                    D &                    P \\
 156 &   0.9 &   0.6 &   1.0 &   0.7 &  -5.4 &   2.4 &    0/I &        104255.40-592456.5 &                    &                    D &                    P \\
 159 &   0.9 &   0.4 &   1.6 &   0.9 &  -6.5 &   2.6 &      A &                           &                    &                    D &                    P \\
 161 &   1.7 &   1.0 &   2.4 &   1.3 &  -5.3 &   2.1 &      A &        104258.65-592534.4 &                    &                    D &                    P \\
 162 &   2.0 &   1.2 &   2.6 &   1.5 & \nodata & \nodata &      A &        104258.78-592503.9 &                    &                    D &                    P \\
 164 &   1.3 &   0.4 &   1.8 &   0.8 & \nodata & \nodata &      A &        104300.29-592526.9 &                    &                    D &                    P \\
\hline
 177 &   1.7 &   0.2 &   2.6 &   0.4 & \nodata & \nodata &      A &        104307.37-592433.6 &                    &                    D &                    Q \\
 179 &   3.6 &   0.0 &  10.1 &   0.9 &  -2.6 &   0.2 &    0/I &        104309.83-592454.2 &                    &                    D &               Q, MSX \\
 183 &   0.7 &   0.5 &   1.2 &   0.8 &  -7.9 &   2.4 &      A &                           &                    &                    D &                    Q \\
 184 &   1.8 &   1.1 &   1.7 &   1.6 &  -5.4 &   2.0 &    0/I &                           &                    &                    D &                    Q \\
 185 &   1.8 &   0.6 &   2.7 &   1.5 &  -5.2 &   2.0 &    0/I &                           &                    &                    D &                    Q \\
 186 &   1.1 &   0.2 &   2.0 &   0.8 & \nodata & \nodata &     II &        104312.06-592458.5 &                    &                    D &                    Q \\
 187 &   1.6 &   1.1 &   1.9 &   1.3 &  -5.4 &   2.0 &    0/I &                           &                    &                    D &                    Q \\
 189 &   0.9 &   0.5 &   1.2 &   0.7 &  -6.3 &   2.5 &    0/I &        104312.59-592516.2 &                    &                    D &                    Q \\
 192 &   1.2 &   0.3 &   2.1 &   0.5 & \nodata & \nodata &     II &        104313.98-592353.0 &                    &                    D &                    Q \\
 194 &   0.8 &   0.6 &   0.8 &   0.9 &  -6.1 &   2.4 &      A &                           &                    &                    D &                    Q \\
 197 &   1.0 &   0.7 &   1.0 &   0.9 &  -5.9 &   2.3 &      A &                           &                    &                    D &                    Q \\
\hline
 372 &   2.0 &   1.2 &   3.2 &   1.0 & \nodata & \nodata &     II &                           &                    &                    D &                    R \\
 391 &   1.0 &   0.1 &   1.1 &   0.4 &  -5.1 &   2.4 &    0/I &                           &                    &                    D &                    R \\
 400 &   2.4 &   1.4 &   3.2 &   1.8 &  -4.4 &   1.6 &    0/I &                           &                    &                    D &                    R \\
 407 &   1.5 &   1.0 &   1.4 &   1.5 &  -5.1 &   2.0 &    0/I &                           &                    &                    D &                    R \\
 416 &   1.3 &   0.8 &   2.1 &   1.0 & \nodata & \nodata &     II &                           &                    &                    D &                    R \\
 417 &   2.2 &   1.3 &   3.3 &   1.4 &  -5.9 &   1.8 &    0/I &                           &                    &                    D &                    R \\
 426 &   1.0 &   0.8 &   1.8 &   0.6 &  -8.8 &   2.8 &     II &                           &                    &                    D &                    R \\
 428 &   0.9 &   0.4 &   1.5 &   0.7 &  -7.0 &   2.7 &     II &        104404.95-590833.6 &                    &                    D &                    R \\
 438 &   2.1 &   0.3 &   2.8 &   1.1 &  -3.8 &   1.5 &    0/I &                           &                    &                    D &                    R \\
 442 &   1.5 &   1.0 &   1.6 &   1.1 &  -6.4 &   2.5 &      A &                           &                    &                    D &                    R \\
\hline
 528 &   1.1 &   0.7 &   2.0 &   0.6 &  -7.2 &   2.8 &     II &        104425.97-593339.0 &                   C5 &                    A &               Cr 232 \\
 556 &   3.6 &   2.0 &   9.0 &   3.2 &  -4.0 &   1.5 &    0/I &        104431.23-593311.2 &                   C5 &                    A &               Cr 232 \\
 563 &   1.7 &   0.0 &   3.7 &   0.2 &  -8.0 &   3.0 &     II &                           &                   C5 &                    A &               Cr 232 \\
 574 &   2.1 &   1.2 &   3.2 &   2.1 &  -4.4 &   1.6 &    0/I &                           &                   C5 &                    A &               Cr 232 \\
 575 &   1.8 &   0.2 &   3.1 &   0.7 &  -7.4 &   2.2 &     II &        104433.23-593348.4 &                   C5 &                    A &               Cr 232 \\
 600 &   1.6 &   0.2 &   2.6 &   0.3 & \nodata & \nodata &     II &                           &                   C5 &                    A &               Cr 232 \\
 612 &   3.4 &   1.9 &   6.4 &   3.0 & \nodata & \nodata &     II &                           &                   C5 &                    A &               Cr 232 \\
 616 &   0.9 &   0.7 &   1.1 &   0.9 &  -6.6 &   2.1 &      A &                           &                    &                    A &               Cr 232 \\
 619 &   1.5 &   1.0 &   1.6 &   1.7 &  -7.2 &   2.2 &      A &        104439.84-593240.2 &                   C5 &                    A &               Cr 232 \\
 621 &   1.4 &   1.0 &   1.6 &   1.2 &  -5.9 &   2.2 &      A &                           &                   C5 &                    A &               Cr 232 \\
 625 &   1.4 &   0.9 &   1.7 &   1.1 &  -6.3 &   2.3 &      A &                           &                   C5 &                    A &               Cr 232 \\
 634 &   1.8 &   1.5 &   2.5 &   1.2 &  -5.6 &   2.2 &      A &                           &                   C5 &                    A &               Cr 232 \\
 642 &   1.9 &   0.8 &   2.8 &   0.9 & \nodata & \nodata &      A &                           &                   C5 &                    A &               Cr 232 \\
 654 &   1.8 &   1.1 &   2.2 &   1.5 &  -5.2 &   2.0 &    0/I &                           &                    &                    A &               Cr 232 \\
 655 &   2.1 &   0.3 &   3.9 &   1.3 &  -5.0 &   2.1 &      A &                           &                    &                    A &               Cr 232 \\
 656 &   2.3 &   1.4 &   3.1 &   1.8 &  -5.2 &   1.9 &      A &                           &                    &                    A &               Cr 232 \\
 657 &   2.2 &   1.3 &   3.3 &   1.6 &  -5.3 &   1.9 &      A &                           &                    &                    A &               Cr 232 \\
 660 &   1.8 &   1.1 &   2.2 &   1.2 &  -8.5 &   2.4 &      A &                           &                   C5 &                    A &               Cr 232 \\
\hline
 699 &   1.0 &   0.8 &   0.7 &   1.0 &  -4.8 &   2.2 &    0/I &        104451.91-602511.9 &                  G18 &                    D &                   3X \\
\hline
1295 &   1.7 &   0.1 &   3.4 &   0.2 & \nodata & \nodata &    III &                           &                    &                    D &                    O \\
1297 &   1.2 &   0.4 &   1.3 &   1.3 &  -5.0 &   2.3 &    0/I &                           &                    &                    D &                    O \\
1302 &   1.4 &   1.1 &   1.9 &   0.6 & \nodata & \nodata &      A &                           &                    &                    D &                    O \\
1304 &   2.6 &   1.5 &   4.2 &   1.5 & \nodata & \nodata &      A &                           &                    &                    D &                    O \\
1308 &   2.0 &   1.2 &   3.1 &   0.9 & \nodata & \nodata &     II &                           &                    &                    D &                    O \\
1312 &   2.2 &   0.1 &   3.7 &   0.4 & \nodata & \nodata &     II &        104748.69-602611.1 &                    &                    D &                    O \\
1313 &   1.3 &   0.8 &   2.8 &   0.6 & \nodata & \nodata &     II &                           &                    &                    D &                    O \\
1315 &   1.5 &   0.4 &   2.1 &   1.2 &  -5.0 &   1.9 &    0/I &        104750.03-602619.1 &                    &                    D &                    O \\
1316 &   1.2 &   0.8 &   1.8 &   1.1 &  -6.3 &   2.5 &      A &                           &                    &                    D &                    O \\
1317 &   2.0 &   1.2 &   1.4 &   1.5 &  -4.6 &   1.9 &    0/I &                           &                    &                    D &                    O \\
1319 &   1.2 &   0.5 &   1.4 &   1.0 &  -4.9 &   1.9 &    0/I &                           &                    &                    D &                    O \\
1320 &   2.6 &   1.5 &   4.3 &   1.5 &  -3.3 &   1.5 &    0/I &                           &                    &                    D &                    O \\
\enddata
\tablecomments{This Table is available in its entirety in a
  machine-readable form in the online journal. 
  A portion is shown here for guidance regarding its form and content
  and to highlight the sources of interest presented in Table
  \ref{obs} and \S\ref{butterflies}.}
\tablenotetext{a}{The Membership columns follow F11.}
\tablenotetext{b}{The Notes column includes the following: O, P, Q, R, or
  Cr 232 refer to membership in the groups 
  discussed in \S\ref{butterflies}; 2X or 3X indicates that the
  \spitzer\ source was matched to 2 or 3 CCCP X-ray sources; and MSX
  means that the YSO is also an \msx\ source.}
\end{deluxetable}


\begin{deluxetable}{@{\hspace{-5mm}}r@{~}c@{\hspace{-5mm}}c@{\hspace{-5mm}}c@{\hspace{-5mm}}c@{\hspace{-5mm}}c@{\hspace{-5mm}}c@{\hspace{-5mm}}c@{\hspace{-5mm}}c@{\hspace{-5mm}}c@{\hspace{-5mm}}c@{\hspace{-5mm}}c}
\tabletypesize{\scriptsize}
\tablecaption{Candidate X-ray-Emitting Stage 0/I YSOs\label{protostars}}
\tablewidth{0pt}
\tablehead{
\colhead{(1)} & \colhead{(2)} & \colhead{(3)} & \colhead{(4)} & \colhead{(5)} & \colhead{(6)} & \colhead{(7)} & \colhead{(8)} & \colhead{(9)} & \colhead{(10)} & \colhead{(11)} & \colhead{(12)} \\[1.2Ex]
\colhead{} & \multicolumn{7}{c}{Physical Parameters from YSO Models} & \colhead{} & \multicolumn{3}{c}{Observed X-ray Properties\tablenotemark{b}}  \\
\cline{2-8} \cline{10-12} \\
\colhead{PCYC} & \colhead{$\langle M_{\star}\rangle$} & \colhead{$\sigma(M_{\star})$} & \colhead{$\langle \log{T_{\star}}\rangle$} & \colhead{$\sigma(\log{T_{\star}})$} & \colhead{$\langle\log{\frac{\dot{M}_{\rm env}}{M_{\star}}}\rangle$} & \colhead{$\langle A_{V,t}\rangle$} & \colhead{Max,$\sigma(A_{V,t})$\tablenotemark{a}} & \colhead{CCCP Match} & \colhead{Net} & \colhead{$E_{\rm median}$} & \colhead{$\log{F_t}$} \\
\colhead{No.} & \colhead{(\Msun)} & \colhead{(\Msun)} & \colhead{(K)} & \colhead{(K)} & \colhead{(yr$^{-1}$)} & \colhead{(mag)} & \colhead{(mag)} & \colhead{CXOGNC J.} & \colhead{Counts} & \colhead{(keV)} & \colhead{(erg/cm$^{2}$/s)}
}
\startdata
\multicolumn{12}{c}{Hard X-ray Counterparts ($E_{\rm median}>2$~keV)\vspace{1.4ex}} \\
\hline
 138 &   3.1 &   0.7 &   3.6 &   0.0 &  -5.0 & $5.0\times 10^2$ & $4.5\times 10^2$ &        104248.10-592539.3 & 101.5 &   3.3 & -13.4 \\
 140 &   0.8 &   0.9 &   3.5 &   0.1 &  -5.8 & $4.1\times 10^0$ & $6.4\times 10^2$ &        104248.50-594835.5 &   5.2 &   2.3 & -14.9 \\
 179 &  10.1 &   0.9 &   3.6 &   0.0 &  -3.6 & $7.2\times 10^2$ & $4.9\times 10^2$ &        104309.83-592454.2 &  44.9 &   4.0 & -13.9 \\
 239 &   1.2 &   1.2 &   3.6 &   0.2 &  -5.6 & $4.6\times 10^1$ & $4.7\times 10^2$ &        104326.70-602622.7 & 149.6 &   4.7 & -12.9 \\
 282 &   1.9 &   1.5 &   3.7 &   0.2 &  -5.6 & $3.1\times 10^1$ & $9.9\times 10^4$ &        104339.83-593300.9 &   3.6 &   3.0 & -14.9 \\
 318 &   2.8 &   1.0 &   3.7 &   0.2 &  -5.8 & $1.5\times 10^1$ & $1.5\times 10^1$ &        104347.79-593304.4 &   5.6 &   2.9 & -14.8 \\
 348 &   3.7 &   2.4 &   3.8 &   0.4 &  -5.4 & $4.9\times 10^1$ & $4.0\times 10^5$ &        104351.67-593257.0 &   9.6 &   2.1 & -14.4 \\
 353 &   1.9 &   1.1 &   3.7 &   0.1 &  -5.8 & $7.0\times 10^0$ & $5.1\times 10^3$ &        104352.89-593921.8 &  92.9 &   2.4 & -13.6 \\
 371 &   4.0 &   1.1 &   3.7 &   0.1 &  -5.4 & $2.5\times 10^1$ & $1.7\times 10^1$ &        104355.73-592417.5 &  55.3 &   2.8 & -13.6 \\
 373 &   3.7 &   1.2 &   3.7 &   0.0 &  -5.3 & $2.9\times 10^1$ & $1.3\times 10^1$ &        104355.80-592421.7 &  11.8 &   3.2 & -14.1 \\
 387 &   1.5 &   1.7 &   3.6 &   0.2 &  -5.4 & $5.7\times 10^1$ & $1.7\times 10^3$ &        104357.77-592833.3 &   8.5 &   2.2 & -14.6 \\
 419 &   2.8 &   1.6 &   3.6 &   0.1 &  -4.8 & $1.8\times 10^2$ & $1.6\times 10^5$ &        104403.08-593748.7 &  19.0 &   2.6 & -14.5 \\
 484 &   9.5 &   0.8 &   4.0 &   0.6 &  -5.2 & $4.1\times 10^1$ & $3.5\times 10^1$ &        104417.86-602745.9 &   5.5 &   4.7 & -14.4 \\
 598 &   2.2 &   1.6 &   3.7 &   0.3 &  -5.3 & $3.7\times 10^1$ & $1.0\times 10^3$ &        104437.14-595802.4 &  29.1 &   4.0 & -13.9 \\
 623 &   3.5 &   1.3 &   3.8 &   0.5 &  -4.9 & $4.1\times 10^1$ & $9.3\times 10^1$ &        104440.31-594339.7 &   2.3 &   5.4 & -14.9 \\
 686 &   0.9 &   1.0 &   3.6 &   0.1 & \nodata & $5.2\times 10^0$ & $1.5\times 10^2$ &        104449.16-595346.6 & 138.3 &   2.9 & -13.5 \\
 699 &   0.7 &   1.0 &   3.5 &   0.1 &  -4.7 & $3.0\times 10^1$ & $1.4\times 10^3$ &        104451.91-602511.9 & 376.8 &   5.2 & -12.3 \\
 713 &   1.6 &   0.7 &   3.7 &   0.1 &  -6.2 & $8.9\times 10^0$ & $7.0\times 10^0$ &        104455.40-592634.5 &  35.5 &   4.4 & -14.1 \\
 774 &   2.1 &   1.1 &   3.7 &   0.1 &  -6.6 & $1.5\times 10^1$ & $1.6\times 10^1$ &        104507.15-600156.9 &  79.2 &   3.0 & -13.8 \\
 897 &   0.6 &   0.5 &   3.6 &   0.1 &  -5.3 & $3.2\times 10^1$ & $7.3\times 10^2$ &        104530.83-595809.5 &   5.3 &   2.2 & -15.0 \\
1005 &   2.4 &   0.9 &   3.7 &   0.1 &  -9.0 & $1.5\times 10^1$ & $8.9\times 10^0$ &        104549.19-601637.3 &  19.4 &   3.0 & -14.3 \\
1073 &   1.7 &   1.0 &   3.6 &   0.0 &  -5.7 & $8.4\times 10^0$ & $4.9\times 10^0$ &        104603.53-595339.3 &  28.6 &   2.2 & -14.3 \\
1185 &   1.0 &   0.9 &   3.7 &   0.2 &  -5.8 & $1.5\times 10^1$ & $2.6\times 10^1$ &        104637.74-594804.4 &  15.9 &   2.2 & -14.6 \\
1236 &   2.7 &   0.4 &   3.6 &   0.0 &  -5.0 & $2.8\times 10^1$ & $6.1\times 10^0$ &        104659.85-602535.0 &  10.3 &   2.3 & -14.4 \\
1315 &   2.1 &   1.2 &   3.6 &   0.0 &  -5.3 & $2.7\times 10^1$ & $1.6\times 10^1$ &        104750.03-602619.1 &  17.2 &   2.7 & -14.0 \\
1367 &   7.6 &   2.5 &   4.0 &   0.2 &  -5.4 & $5.6\times 10^2$ & $1.7\times 10^6$ &        104824.37-600759.5 &  22.0 &   4.2 & -13.9 \\
\hline
\vspace{-0.9ex} \\
\multicolumn{12}{c}{Soft X-ray Counterparts ($E_{\rm median}\le 2$~keV)\vspace{1.4ex}} \\
\hline 
  83 &   0.8 &   0.6 &   3.6 &   0.0 &  -5.1 & $2.1\times 10^1$ & $2.9\times 10^1$ &        104216.00-593607.6 &  13.5 &   1.5 & -14.4 \\
 148 &   0.6 &   0.5 &   3.6 &   0.1 &  -4.5 & $4.6\times 10^1$ & $1.4\times 10^2$ &        104251.41-592535.6 &   5.5 &   1.5 & -14.9 \\
 156 &   1.0 &   0.7 &   3.7 &   0.2 &  -5.4 & $2.8\times 10^1$ & $2.9\times 10^2$ &        104255.40-592456.5 &   3.6 &   1.5 & -14.9 \\
 189 &   1.2 &   0.7 &   3.6 &   0.1 &  -6.3 & $1.0\times 10^1$ & $3.3\times 10^1$ &        104312.59-592516.2 &  30.1 &   1.8 & -14.4 \\
 288 &   2.6 &   1.0 &   3.6 &   0.0 &  -5.1 & $2.5\times 10^2$ & $5.4\times 10^3$ &        104341.07-594537.2 &  51.1 &   1.5 & -14.4 \\
 346 &   2.5 &   0.6 &   3.6 &   0.0 &  -4.7 & $3.1\times 10^2$ & $6.6\times 10^2$ &        104351.59-600525.2 &  17.0 &   1.3 & -14.6 \\
 352 &   1.7 &   1.6 &   3.6 &   0.2 &  -5.7 & $2.3\times 10^1$ & $6.5\times 10^4$ &        104352.58-593025.1 &   8.7 &   1.4 & -15.1 \\
 355 &   2.0 &   0.9 &   3.7 &   0.1 &  -5.3 & $3.2\times 10^1$ & $5.0\times 10^4$ &        104353.56-593410.1 &   5.6 &   1.5 & -15.0 \\
 369 &   2.1 &   1.5 &   3.7 &   0.1 &  -6.3 & $4.4\times 10^0$ & $1.0\times 10^3$ &        104355.20-600519.8 &  11.5 &   1.2 & -14.4 \\
 376 &   0.5 &   0.7 &   3.6 &   0.3 & \nodata & $4.8\times 10^0$ & $4.2\times 10^1$ &        104355.92-592537.6 &  19.5 &   1.5 & -14.7 \\
 424 &   2.8 &   1.1 &   3.7 &   0.2 &  -5.5 & $5.6\times 10^0$ & $5.5\times 10^0$ &        104403.87-593308.3 &   4.7 &   1.0 & -15.4 \\
 432 &   2.3 &   0.9 &   3.7 &   0.0 &  -5.3 & $1.2\times 10^2$ & $2.0\times 10^2$ &        104406.04-595510.6 &  11.1 &   1.7 & -14.6 \\
 450 &   2.3 &   1.2 &   3.7 &   0.2 &  -5.7 & $2.8\times 10^0$ & $1.8\times 10^1$ &        104409.95-593436.1 &  27.2 &   1.5 & -14.3 \\
 459 &   1.3 &   0.7 &   3.6 &   0.1 &  -7.2 & $8.7\times 10^0$ & $5.4\times 10^1$ &        104411.75-592431.1 &  19.5 &   1.6 & -14.5 \\
 496 &   1.7 &   1.4 &   3.6 &   0.1 &  -5.3 & $4.5\times 10^1$ & $1.6\times 10^2$ &        104421.14-593000.9 &   4.6 &   1.3 & -15.1 \\
 505 &   0.8 &   1.0 &   3.6 &   0.1 &  -6.2 & $3.4\times 10^0$ & $3.1\times 10^2$ &        104423.08-600819.3 &  26.9 &   1.2 & -14.4 \\
 508 &   0.7 &   0.9 &   3.6 &   0.2 &  -5.8 & $5.8\times 10^0$ & $6.3\times 10^2$ &        104423.32-592513.5 &   5.5 &   1.6 & -15.1 \\
 521 &   1.3 &   1.1 &   3.7 &   0.3 &  -8.3 & $5.1\times 10^0$ & $2.0\times 10^3$ &        104425.68-593923.5 &  55.1 &   1.3 & -14.5 \\
 548 &   1.7 &   1.7 &   3.6 &   0.1 &  -5.4 & $5.7\times 10^1$ & $6.7\times 10^2$ &        104429.88-594533.8 &   4.6 &   1.3 & -15.2 \\
 556 &   9.0 &   3.2 &   4.0 &   0.6 &  -4.9 & $8.7\times 10^1$ & $2.4\times 10^3$ &        104431.23-593311.2 &  18.2 &   1.9 & -14.3 \\
 593 &   1.5 &   1.2 &   3.7 &   0.2 &  -7.7 & $5.6\times 10^0$ & $1.1\times 10^4$ &        104436.06-594258.4 &  12.9 &   1.5 & -14.7 \\
 667 &   1.5 &   0.8 &   3.6 &   0.0 &  -5.4 & $3.7\times 10^1$ & $1.3\times 10^3$ &        104446.07-595816.8 &  12.6 &   1.5 & -14.9 \\
 694 &   1.5 &   1.5 &   3.6 &   0.2 &  -5.5 & $4.0\times 10^1$ & $1.3\times 10^5$ &        104450.81-594442.9 &  58.2 &   1.6 & -14.2 \\
 768 &   2.6 &   1.4 &   3.7 &   0.1 &  -5.9 & $1.1\times 10^2$ & $4.6\times 10^2$ &        104505.75-594405.0 &  16.4 &   1.3 & -14.7 \\
 795 &   1.3 &   0.9 &   3.7 &   0.2 &  -6.2 & $4.3\times 10^0$ & $2.4\times 10^0$ &        104511.09-594533.5 &  66.4 &   1.4 & -14.1 \\
 890 &   0.7 &   0.9 &   3.6 &   0.2 &  -5.8 & $5.9\times 10^0$ & $1.4\times 10^3$ &        104529.17-601357.0 &  20.7 &   1.4 & -14.5 \\
 928 &   1.0 &   0.8 &   3.6 &   0.1 &  -6.0 & $9.7\times 10^0$ & $3.0\times 10^2$ &        104536.94-595611.3 &   9.5 &   2.0 & -14.6 \\
 987 &   0.6 &   0.7 &   3.5 &   0.2 &  -5.3 & $3.3\times 10^1$ & $9.1\times 10^2$ &        104546.73-600343.3 &  14.7 &   0.9 & -15.1 \\
\phn1025\tablenotemark{c} &   4.2 &   0.5 &   3.7 &   0.0 &  -5.3 & $5.1\times 10^1$ & $6.2\times 10^1$ &        104553.71-595703.9 &  78.1 &   1.5 & -13.8 \\
1055 &   3.6 &   2.8 &   4.0 &   0.6 &  -5.2 & $7.7\times 10^0$ & $5.1\times 10^0$ &        104559.44-600517.9 & 352.0 &   1.6 & -13.2 \\
1141 &   2.5 &   0.8 &   3.7 &   0.1 &  -5.8 & $1.2\times 10^1$ & $1.8\times 10^1$ &        104622.54-594523.9 &  14.2 &   1.7 & -14.3 \\
1193 &   0.5 &   0.7 &   3.5 &   0.2 &  -5.8 & $4.0\times 10^0$ & $1.0\times 10^3$ &        104642.52-600826.7 &   7.9 &   1.5 & -14.9 \\
1204 &   1.3 &   0.7 &   3.6 &   0.1 &  -5.5 & $4.4\times 10^1$ & $1.8\times 10^3$ &        104646.68-600932.3 &   4.0 &   1.4 & -14.9 \\
1276 &   0.8 &   0.8 &   3.6 &   0.1 &  -8.1 & $7.7\times 10^0$ & $2.8\times 10^1$ &        104717.42-600906.1 &  61.0 &   1.6 & -13.9 \\
1390 &   0.4 &   0.6 &   3.5 &   0.1 &  -5.9 & $4.0\times 10^0$ & $2.6\times 10^2$ &        104836.21-601408.7 &   7.7 &   1.3 & -14.9 \\
\enddata
\tablenotetext{a}{The probability distributions of $A_{V,t}$ can be strongly skewed,
  with long tails to high values. This often yields very large values
  for $\sigma(A_{V,t})$, hence in cases where $\sigma(A_{V,t})\gg
  \langle A_{V,t}\rangle$ we define Max$(A_{V,t})\equiv \sigma(A_{V,t})$.
}
\tablenotetext{b}{Columns (10)--(12) list the quantities
  NetCounts$_{\rm t}$, EnergyFlux$_{\rm t}$, and MedianEnergy$_{\rm
      t}$ from Table 1 of \citet{CCCPcatalog}. 
}
\tablenotetext{c}{PCYC 1025 is located in the the very crowded center
  of the Treasure Chest, hence it is very likely a combination of
  multiple confused sources in the IRAC beam, and the physical
  association of the CCCP source with the IRAC source is highly questionable.}
\end{deluxetable}

\end{document}